\documentclass[12pt]{iopart}

\expandafter\let\csname equation*\endcsname\relax  
\expandafter\let\csname endequation*\endcsname\relax   

\usepackage{physics}  

\usepackage{iopams}  
\usepackage{graphicx} 
\usepackage{subcaption}

\begin{document}

\title{Exploring Dielectric Properties in Models of Amorphous Boron Nitride}
\author{Thomas Galvani$^{1}$, Ali K. Hamze$^{2}$, Laura Caputo$^{3}$, Onurcan Kaya$^{1,4}$, Simon Dubois$^{3}$, Luigi Colombo$^{6}$,  Viet-Hung Nguyen$^{3}$, Yongwoo Shin$^{2}$, Hyeon-Jin Shin$^{5}$, Jean-Christophe Charlier$^{3}$ and Stephan Roche$^{1,7}$}
\address{$^1$ Catalan Institute of Nanoscience and Nanotechnology (ICN2), CSIC and BIST, Campus UAB, Bellaterra, 08193, Barcelona, Spain}
\address{$^2$ Advanced Materials Lab, Samsung Semiconductor Inc., Cambridge, MA 02138, USA}
\address{$^3$ Institute of Condensed Matter and Nanosciences, Universit\'e catholique de Louvain (UCLouvain), Chemin des Étoiles 8, Louvain-la-Neuve 1348, Belgium}
\address{$^4$ School of Engineering, RMIT University, Melbourne, Victoria, 3001, Australia }
\address{$^5$ Department of Semiconductor Engineering, School of Electrical Engineering and Computer Science, Gwangju Institute of Science and Technology (GIST), Gwangju 61005, Republic of Korea}
\address{$^6$ CNMC, LLC, Dallas, TX, 75248, USA}
\address{$^7$ ICREA Institucio Catalana de Recerca i Estudis Avancats, 08010 Barcelona, Spain}
\ead{stephan.roche@icn2.cat}

\begin{abstract}

We report a theoretical study of dielectric properties of models of amorphous Boron Nitride, using interatomic potentials generated by machine learning. We first perform first-principles simulations on small (about $100$ atoms in the periodic cell) sample sizes to explore the emergence of mid-gap states and its correlation with structural features. Next, by using a simplified tight-binding electronic model, we analyse the dielectric functions for complex three dimensional models (containing about $10.000$ atoms) embedding varying concentrations of ${\rm sp^{1}, sp^{2}}$ and ${\rm sp^3}$ bonds between B and N atoms. Within the limits of these methodologies, the resulting value of the zero-frequency dielectric constant is shown to be influenced by the population density of such mid-gap states and their localization characteristics. We observe nontrivial correlations between the structure-induced electronic fluctuations and the resulting dielectric constant values. Our findings are however just a first step in the quest of accessing fully accurate dielectric properties of as-grown amorphous BN of relevance for interconnect technologies and beyond.
\end{abstract}

\section{Introduction}

As the need for larger information storage and processing explodes, improvements of computing device performances such as the lateral scaling, and integration of active devices in the back-end-of-line (BEOL) are crucially needed \cite{TENG2023112086}. In particular, the optimization of interconnect technologies becomes increasingly important as the signal delay of the interconnect increases rapidly compared to gate delay of the transistor \cite{STREITER1997429,Chaudhry_2013}. It is therefore critically important to develop new materials or improve existing ones to decrease interconnect energy loss, through dielectric constant and metal resistivity reduction (resistance-capacitance (RC) delay). Since the early 2000s, SiCOH, with a dielectric constant of about 3.0, has been the material of choice instead of $\mathrm{SiO_2}$, which has a dielectric constant of 4.0. Then, by embedding pore structures with a dielectric constant of 1.0, the dielectric constant of porous-SiCOH (p-SiCOH) could be linearly reduced, although unfortunately this decreases its Young's modulus exponentially, making it difficult to perform post deposition processes such as chemical mechanical processing (CMP) and packaging ports. Ultimately, this effectively limits the dielectric constant of p-SiCOH in such applications to about 2.5 \cite{C3TC00587A,Grill2016}.

Recent reports of amorphous Boron Nitride (aBN) have shown unprecedented ultralow dielectric constant of 1.8, together with a robust Young's modulus of over 50 GPa in absence of pore morphology, sparking a great interest in view of the long sought-after improved interconnects technologies \cite{Hong2020,Lin2022}. Additionally, aBN has been found to display good mechanical properties overall as well as excellent thermal properties and diffusion barrier against metal migration. All such properties turn out to be perfectly suited for the development of next-generation interconnect technologies \cite{Grill2016, Noguchi2005, Palov2106}.

Here, within this context, we present a theoretical study of electronic and dielectric properties of complex models of aBN, and we attempt to correlate them with the system's atomic structure. Indeed, because aBN is an amorphous system, its structure, and therefore its properties of merit, depend strongly on its fabrication parameters (growth rate, temperature, B-N stoichiometry,...) \cite{Lin2022, Chen2023arxiv}. To progress in material optimization, proper and accurate simulation studies are thus highly desirable.

However, unfortunately, the modeling of {\it highly disordered (large-scale) materials} is generally a very challenging task for predictive theoretical investigation. Indeed, realistic modeling of nanoscale electronic properties in such complex systems benefits strongly from the predictive power of first-principles techniques, which can be quite accurate. But at the same time, such methods are also limited in accessible sample sizes (typically several thousands of atoms) and are generally unable to capture disorder effects such as multiple scattering, impurity states, quantum interference and localization in large-scale (realistic) models.

To cope with the complexity of the modeling problem, we employ a two-pronged approach. 
First, we generate a dataset of small aBN samples (lattices with $\sim 10^2$ atoms in their periodic cell) through simulated annealing with machine learned force fields \cite{Kaya2023} and compute their electronic and dielectric properties using DFT. This allows us to explore the influence of the system's microstructure on its electronic and dielectric properties with \textit{ab initio} precision. In this framework, we show the emergence of some correlations between the nature of the structural disorder and bonding character statistics inside the sample, and the resulting formation of mid-gap states and behavior of the dielectric constant (Section \ref{s:SmallDFT}).
Second, we construct large scale atomic structural models ($\sim 10^4$ atoms) using Molecular Dynamics with such machine learning potentials, which enjoy {\it ab-initio} accuracy \cite{Kaya2023,kaya2023impact}, and evaluate the electronic and dielectric properties of these large systems through a simple Slater-Koster tight-binding model. Although this latter model presents significant limitations, it allows us to get a first picture of the electronic and dielectric properties in large, highly complex three-dimensional geometries mixing Boron and Nitrogen atoms, linked together by ${\rm sp^{1},sp^{2}}$ and ${\rm sp^3}$ bonds. The obtained results are not aimed at quantitative predictions but we expect a certain validity concerning some qualitative trends.

Finally, by summarizing the limitations of employed methodologies, we will put in perspective the obtained results and discuss the need for more sophisticated modeling strategies to achieve better future quantitative predictions, able to guide experiments and research at the industrial level.

\section{First-principles simulations of small size models }
\label{s:SmallDFT}

\subsection{Generation of a dataset of small structures}
\label{ss:datasetGeneration}
Calculations of the dielectric constant of the small (i.e., typically $100$ atoms in the unit cell) structures were done with the Vienna \emph{ab initio} Simulation Package (VASP) and projector-augmented wave pseudopotentials \cite{vasp1,vasp2,vasp3,vasp4, vasp5} using density functional perturbation theory \cite{vasp6}. The exchange-correlation functional was calculated in the generalized gradient approximation  \cite{gga-pbe}, and the $1s$ electrons were treated as core states for both B and N atoms. We only consider the electronic contribution to the dielectric constant and only calculate the static value. For all calculations, we use a cutoff energy of 520 eV in the plane-wave expansion and converge the energy to $10^{-6}$ eV. A $k$-point grid density of 100 $k$-points per $\mathrm{\AA}^{-3}$ of the reciprocal cell was used for almost all calculations in this dataset, with only 1 data point with 80 $k$-points per $\mathrm{\AA}^{-3}$ of the reciprocal cell. With these calculation parameters, the band gap of hexagonal BN is 4.48 eV, while the dielectric constants are $\left(\epsilon_{xx}=\epsilon_{yy}, \epsilon_{zz}\right) = (4.58, 2.15)$.

To reduce the computational cost and speed up the generation of the amorphous structures, we use two machine-learned force fields. The first is a Gaussian approximation potential trained by Kaya \emph{et al}. \cite{Kaya2023} which was used for all structures with B/N ratios not equal to 1, as well as some of the structures with B/N = 1. Since the potential of Kaya \emph{et al}. \cite{Kaya2023} was trained using first-principles data generated with Quantum Espresso, for consistency, the final structures were fully relaxed with VASP (until the atomic forces were $\leq 10^{-3}\ \mathrm{eV/\AA}$) before the dielectric constants were calculated. 
To avoid the need for structure optimization using VASP for later structures, a second force field was trained \cite{hamze_shin} on first principles data generated from VASP using DeePMD \cite{deepmd1-WANG2018178, deepmd2-10.1063/5.0155600, deepmd3-LU2021107624, deepmd4-doi:10.1021/acs.jctc.2c00102} and DP-GEN \cite{dpgen-PhysRevMaterials.3.023804}. The training data for the DeePMD potential was generated using the same pseudopotentials and exchange-correlation functional approximation as used for the dielectric constant calculations (detailed above). The model was trained on 181621 structures with 64, 96, and 100 atoms, and validated on 20181 structures. The mean absolute error of the energy per atom predicted by the force field on the validation set was 27.3 meV/atom, and the mean absolute error of the force components was 0.355 eV/$\mathrm{\AA}$.
Structures generated with this potential all had B/N = 1 and did not require additional relaxation before the dielectric constant was calculated. 

The same procedure was followed for the generation of the small structures regardless of the potential used. First, B and N atoms were randomly scattered using PACKMOL \cite{packmol-https://doi.org/10.1002/jcc.21224} in a box with volume chosen to give a target initial density for the structure at hand. Densities were centered around 2.1 g/cm$^{-3}$ (the density of hexagonal BN and approximately that of experimentally grown low-$k$ a-BN \cite{Hong2020, Lin2022}), 3.46 g/cm$^{-3}$ (the density of cubic BN), and 2.8 g/cm$^{-3}$ (in between). Simulated annealing was used to generate low-energy structures from the initial randomly scattered atoms. Each structure is heated from 300 K to 3000 K over 200 ps of simulation time, held at 3000 K for 20 ps, and then cooled back to 300 K at either 100 K/ps or 13.5 K/ps before a final relaxation. Some structures were generated in the NPT ensemble, and some were generated in the NVT ensemble to keep the density fixed. When the NPT ensemble was used, the lattice in the final relaxation was also allowed to relax, and when it wasn't, the lattice was held fixed.

\subsection{Dataset exploration}

Using the 209 aBN samples in the dataset, we try to reveal the relationship between their microstructure and their static dielectric constant $\epsilon_1$. We first investigate the effect of B/N imbalance, density, and hybridization of atoms (${\rm sp^1, sp^2}$, and ${\rm sp^3}$) on the dielectric constant and electronic band gap. Then, we use the Cowley short-range order parameter (SRO) \cite{Cowley1950} as a way to evaluate the effects of disorder in Boron-Nitrogen alternation.

\begin{figure}
 \begin{subfigure}{0.49\textwidth}
     \includegraphics[width=\textwidth]{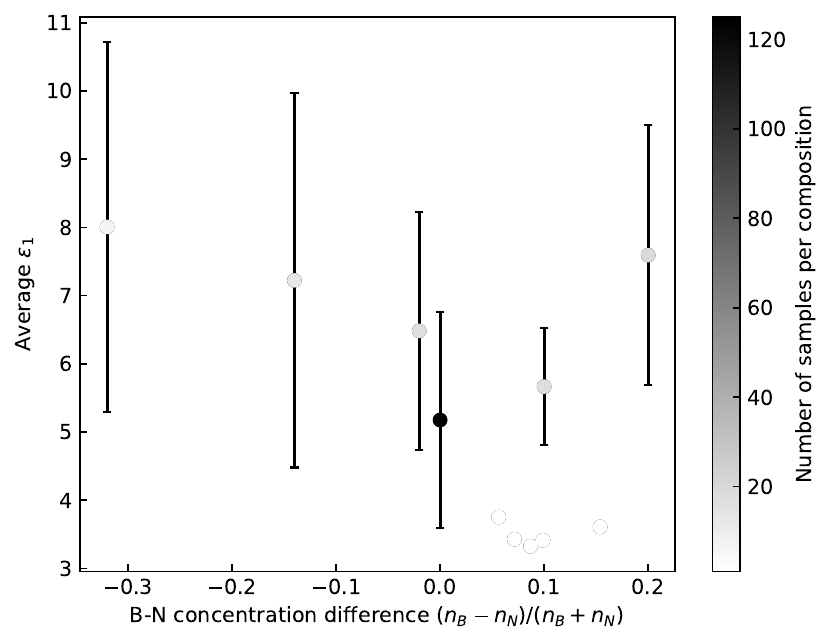}
     \caption{ }
     \label{fig:BN_bias}
 \end{subfigure}
 \begin{subfigure}{0.49\textwidth}
     \includegraphics[width=\textwidth]{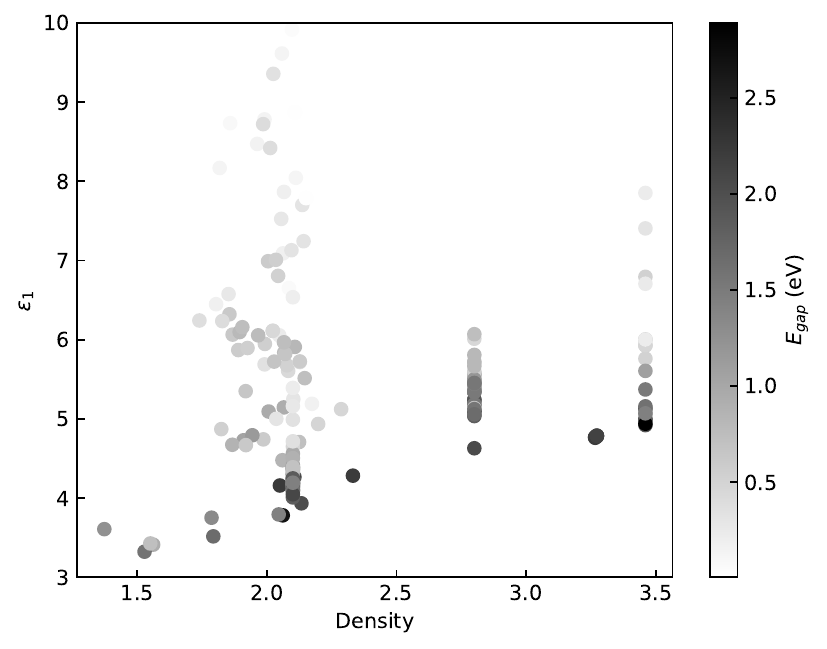}
     \caption{ }
     \label{fig:CN_eps}
 \end{subfigure}
 \begin{subfigure}{\textwidth}
  \includegraphics[width=1\textwidth]{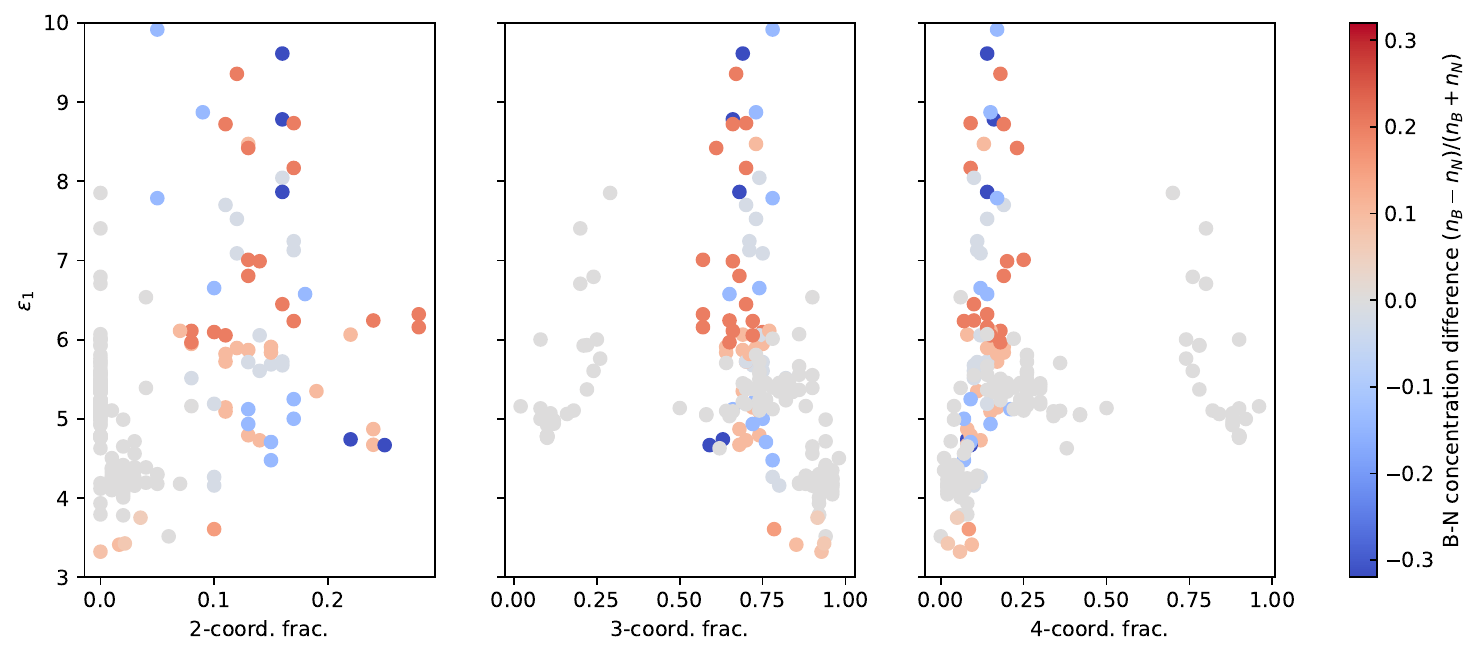}
    \centering
     \caption{ }
     \label{fig:Density_eps}
 \end{subfigure}
 \caption{Distribution of dielectric constant, $\epsilon_1$, of small aBN samples with various B/N compositions (a), densities (b) and different hybridizations of atoms (c). In panel (a), average dielectric constants were estimated by averaging over all dataset samples with a given B-N composition, while error bars represent the associated standard deviation. Sixteen out of the 209 samples were found to exhibit high dielectric constants $\epsilon_1>10$ and are excluded from the above plots for lisibility. Out of these sixteen samples, seven have $\epsilon_1>20$ and up to a few hundreds, and these outliers are excluded from the averages in panel (a).}
 \label{fig:BNDensityCN}
\end{figure}

Let us first examine the question of B-N imbalance. Figure \ref{fig:BNDensityCN}-a displays the average dielectric constant of dataset samples with a given difference in their Boron and Nitrogen concentrations $\Delta c = \frac{n_B - n_N}{n_B + n_N}$.\footnote{We prefer this metric of B-N imbalance over the B/N ratio because it is symmetric in B and N.} In agreement with Lin and co-authors \cite{Lin2022}, we find that B-N imbalance leads to a significant increase in the dielectric constant, both in the B-rich and N-rich cases. 

Figure \ref{fig:BNDensityCN}-b displays the observed relationship between density, dielectric constant and band gap. 
We recover the usual trends that a lower density and a higher band gap favor a lower dielectric constant (as can be intuited from equation \ref{eq:epsilonReal} below, ($\epsilon_1-1) \propto \Omega^{-1}$ where $\Omega$ is the cell volume, and the presence of transition energies $E_c-E_v$ in the denominators), but we attract the reader's attention to the strong variability in $\epsilon_1$ even at fixed density. This variability, in turn, is partially explained by the variation of the gap (color on figure \ref{fig:BNDensityCN}-b), which is an electronic property related to the precise details of the system's microstructure.

\begin{figure}[ht]
    \centering
    \includegraphics[width=1.0\columnwidth]{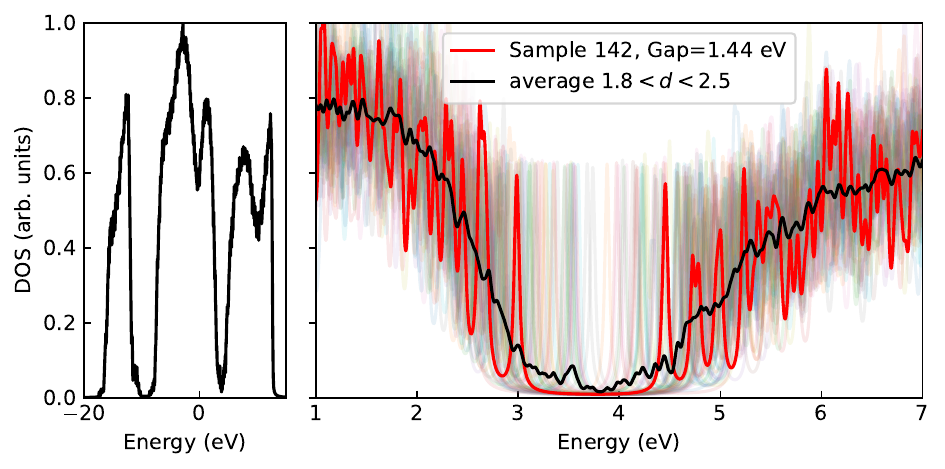}
    \caption{Left panel: averaged DOS of low-density, equal B and N concentration aBN samples (densities comparable to hBN - 48 samples). Right panel: zoom on the region of the electronic gap for these samples. The red curve depicts one of the samples averaged over, for reference, while the DOS of other samples of that set are represented as translucent curves. Individual DOS are computed from the Kohn-Sham energies with a Lorentzian broadening of $26 \text{ meV}$.}
    \label{fig:aDOS}
\end{figure}

To provide insight on these electronic properties, we display in figure \ref{fig:aDOS} the ensemble averaged Density of States (DOS) over all equal concentration ($\Delta c = 0$) low-density samples in the dataset, i.e. samples with a density around that of hBN or lower, as well as one representative sample of this set.\footnote{We choose this set of samples because they are of comparable density with that of hBN.} As can be seen on figures \ref{fig:BNDensityCN}-b and \ref{fig:aDOS}, typical gaps for individual samples are found to vary between $0$ and $\sim 3 \text{ eV}$, significantly lower than the DFT gap of hBN. The DOS of the representative sample visibly displays mid-gap states, which in these small samples effectively set the value of the electronic gap. Given the data of figure \ref{fig:BNDensityCN}-b, this suggests that these mid-gap states may have a notable influence on the system's dielectric constant. The exact energy of these states depends on the specific structure of the sample, so that in a larger sample displaying many different local atomic configurations, one may expect to see a distribution of these states inside what would otherwise be the gap of a ``reference crystalline structure", in a manner akin to the tail states typically observed in other amorphous solids \cite{DRABOLD1998153}. To test this idea, we have performed an ensemble average of the DOS in the subset of the database described above. From this, it can indeed be seen that these mid-gap states effectively fill the gap. In disordered systems, even in this case, such states often do not actually contribute to DC conduction due to localization effects \cite{DRABOLD1998153}. However, likely since the samples of the dataset are small (a few hundreds of atoms), we could not convincingly demonstrate localization in this particular case. We will discuss this in Section \ref{s:TightBinding}, where we consider larger samples using a simple tight-binding model. We still mention here that, even if these states \textit{are} localized, they can still contribute to the dielectric constant of the system by inducing low-transitions to or from delocalized states (see Section \ref{s:TightBinding} for details). 

\begin{figure}[ht]
    \centering
    \includegraphics[width=0.6\columnwidth]{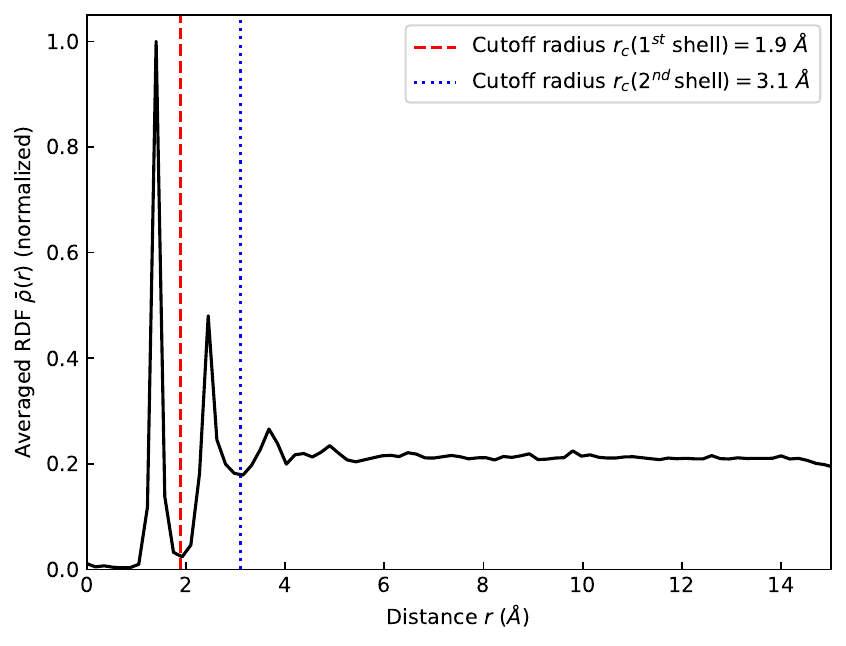}
    \caption{Radial distribution function for small aBN samples averaged over the whole dataset. A marked peak corresponding to the first nearest neighbor shell can be observed, which allows the definition of a cutoff radius for first-nearest neighbor interaction at $r_c=1.9 \text{ \AA}$. A weaker second peak is also visible, which allows for the estimation of a second nearest neighbor shell cutoff $r_c^{(2)}=3.1 \text{ \AA}$.}
    \label{fig:aRDF}
\end{figure}

To get a better understanding of the samples' microstructures, we examine their Radial Distribution Functions (RDF). Figure \ref{fig:aRDF} depicts an ensemble average for the RDF of all samples in the dataset. There is a very clear first peak, allowing us to define a first nearest neighbor shell through a cutoff radius of $r_c=1.9 \text{ \AA}$. 

Using this definition for the first nearest neighbor shell, we display in figure \ref{fig:BNDensityCN}-c the correlations between the coordination number fractions of the samples and their dielectric constants. We define the coordination number of a given atom by counting its neighbors within the cutoff radius $r_c$, with the idea that $2/3/4$-coordinated atoms typically correspond to $\mathrm{sp}^1/\mathrm{sp}^2/\mathrm{sp}^3$ environments. The coordination number fractions $f_i$ are then the ratio of the number $n_i$ of $i$-coordinated atoms in a sample over its total number of atoms $n_{tot}$: $f_i=n_i/n_{tot}$. In this picture, structures with $f_3\to1$ tend to be ordered $\mathrm{sp}^2$ dominated structures, close to ``hBN order", while structures with $f_4\to1$ tend to be close to an ordered 3D allotrope of BN. Let us first focus on the second panel of figure \ref{fig:BNDensityCN}-c, depicting the fraction $f_3$ of $\rm sp^{2}$ coordinated atoms. Samples with the lowest dielectric constant $\epsilon_1$ occur for $f_3=1$, i.e. fully $\rm sp^{2}$-coordinated samples. It can be seen that, as $f_3$ decreases away from 1, $\epsilon_1$ tends to increase steadily: in effect, non-$\rm sp^{2}$ coordinated atoms in an $\rm sp^{2}$-rich structure appear to act as defects which degrade the dielectric properties. The third panel of figure \ref{fig:BNDensityCN}-c shows the same behavior for $\rm sp^{3}$-rich structures, although the latter are seen to have a higher dielectric constant compared to the $\rm sp^{2}$-rich case. This last point can likely be understood from the fact that the $\rm sp^{3}$-dominated structures tend to be markedly denser, as expected structurally. In fact, since the samples in the dataset mostly have $\rm sp^{2}$ and $\rm sp^{3}$ coordinated atoms (with a small fraction of $\rm sp^{1}$), $f_3+f_4\approx 1$ and so the second and third panels are effectively mirrors of each other. As can be seen in the first panel, since no structures are $\rm sp^{1}$-dominated, $\rm sp^{1}$-coordinated atoms tend to always act as defects detrimental to $\epsilon_1$. We should note, however, that such $\rm sp^{1}$ bonds are likely very reactive, and that in real samples we may observe different effects due to contaminants which are not modeled here.

To further this discussion on the local bonding character in the samples, we now take the view of aBN as a disordered binary (A-B) alloy and examine the Boron-Nitrogen alternation in the samples. To this end, we rely on the Cowley Short Range Order (SRO) \cite{Cowley1950}. The SRO for the $i^{th}$ nearest neighbors shell of a given atom $j$ of type $B$ is defined as follows:
\begin{equation}
\label{eq:SRO}
{\rm SRO}_{B_j}(i)=1-\frac{1}{c_A}\frac{n_i^{(j)}(A)}{N_i^{(j)}}
\end{equation}
where $n_i(A)$ is the number of atoms of type $A$ in atom $j$'s $i^{th}$ shell, $N_i^{(j)}$ the total number of atoms in that same shell, and $c_A$ is the global concentration of $A$ in the sample. The ratio $n_i^{(j)}(A)/N_i^{(j)}$ can be seen as the local concentration of A atoms in the $i^{th}$ shell: if the attribution of atomic types in the alloy were random, then on average one would have $n_i^{(j)}(A)/N_i^{(j)}=c_A$ and therefore the SRO would vanish for all shells. Oppositely, if the alloy was perfectly alternating in the sense that consecutive shells alternate in composition as purely one type of atoms (i.e. if $j$ is a B atom, its first shell contains only A atoms, its second shell only B atoms, etc.), then the SRO would oscillate between $1-\frac{1}{c_A}$ ($-1$ for equal concentrations alloys) and $1$ for odd and even shells respectively. The average SRO of order $i$ relative to a given atomic type (say $B$) for a sample is then obtained by averaging over all atoms of that type:
\begin{equation}
{\rm SRO}_{B}(i)=\sum_j {\rm SRO}_{B_j}(i)
\end{equation}
It remains to be seen how nearest neighbor shells are to be attributed in aBN, given the variability in the samples. To this end, we follow (and use) the Pyscal \cite{Menon2019} implementation of the Cowley SRO. Using the averaged RDF displayed in figure \ref{fig:aRDF}, we define a cutoff for the $2^{nd}$ nearest neighbor shell, $r_c^{(2)}=3.1 \text{ \AA}$. Then, for each atom $j$, the average distance between its closest and furthest neighbors within $r_c^{(2)}$ is used to define the boundary between the first and second shells for the purpose of the SRO calculation.

\begin{figure}
 \begin{subfigure}{0.49\textwidth}
   \includegraphics[width=\textwidth]{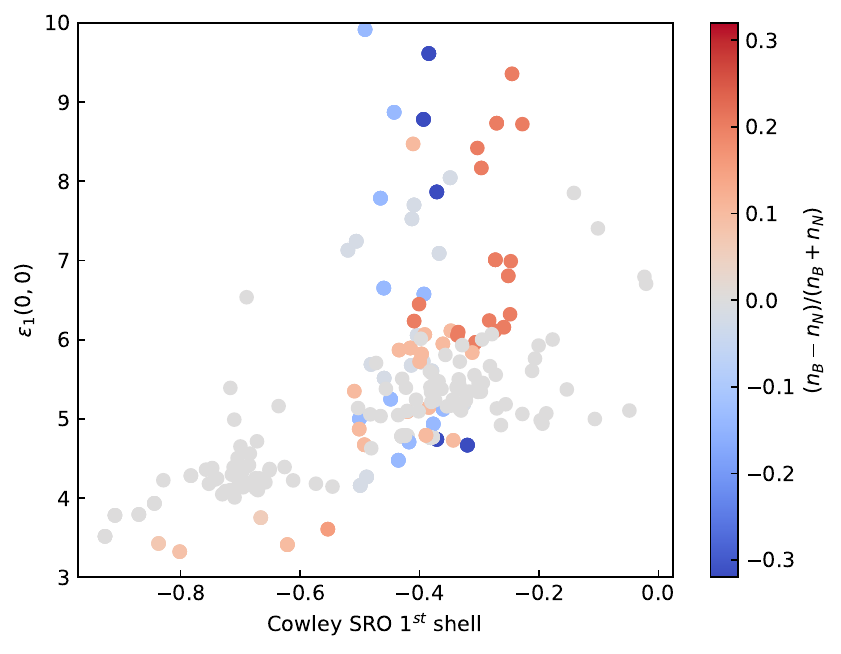}
     \caption{ }
 \end{subfigure}
 \hfill
 \begin{subfigure}{0.49\textwidth}
   \includegraphics[width=\textwidth]{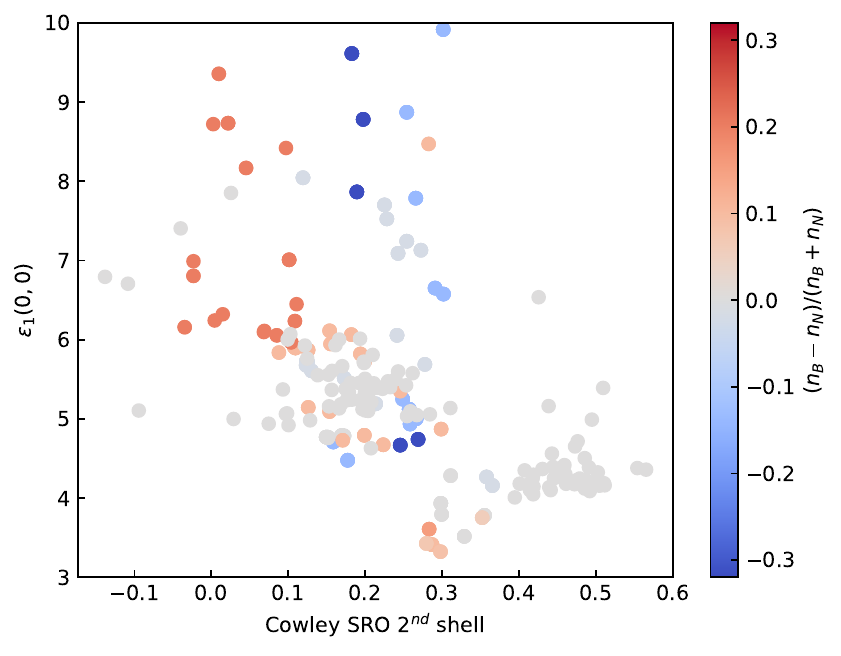}
     \caption{ }
 \end{subfigure}
 \caption{Short range order of aBN samples based on first (a) and second nearest neighbor shells (b).}
 \label{fig:SRO}
 \end{figure}

Figure \ref{fig:SRO} shows the average SRO relative to Boron atoms for the first and second shells for all structures in the dataset. Overall, we observe that structures with the lowest dielectric constants tend to have the lowest first-shell SRO, which, as per the previous discussion, corresponds to the more ordered structures in terms of alternation. A similar trend can be observed from the second shell SRO, for which the highest values correspond to the more ordered alternation. 
It has been suggested by several authors \cite{Glavin2016,Lin2022} that Boron clusters or Boron-Boron bonds could lead to a high dielectric constant. Glavin and co-authors \cite{Glavin2016} have specifically noted the possibility that B-B bonds could create mid-gap states and conductive pathways in their samples. This is consistent with our analysis thus far, and echoes with the idea that non-alternating samples in our dataset also tend to exhibit higher dielectric constants.

\subsection{Partial conclusions}
\label{ss:SmallSamplesConclusion}

To summarize, we have so far observed that departure from an $\rm sp^{2}$- or $\rm sp^{3}$-rich phase tended to produce samples with comparatively higher dielectric constants compared to the ``pure" structure (with $\rm sp^{2}$-rich phases producing the lowest dielectric constants in general), as did departing from a B-N alternating structure in the Cowley SRO sense (with equal B and N concentrations and alternating samples producing the lowest dielectric constants in general). Both of these indicators can be seen as measures of disorder in the samples. It thus follows that, with the important caveat that only small unit cells were considered in the dataset, more disordered aBN samples are expected to display higher dielectric constants. This is in agreement with the findings of Lin and coauthors \cite{Lin2022}, who found that slower growth rates led to films with lower dielectric constants, since slower growth rates tend to produce samples with less disorder.

However, taking this line of reasoning to the extreme may suggest that the systems with the lowest dielectric constants would be the most ordered ones, i.e. the crystals. This seems paradoxical, as many works \cite{Abbas2018, Hong2020, Lin2022} have shown that aBN can have a lower dielectric constant than its crystalline counterparts. For this reason, we turn to the investigation of large samples.

\section{Molecular dynamics and tight-binding investigation of large size models}
\label{s:TightBinding}

We now move to the study of larger systems, using a combination of molecular dynamics to generate large structural atomic models, and a simple tight-binding model to investigate the electronic and dielectric response of these structures.

\subsection{Generation of large structural atomic models}\label{s:StructuralModels}

\begin{figure}[ht]
    \includegraphics[width=\columnwidth]{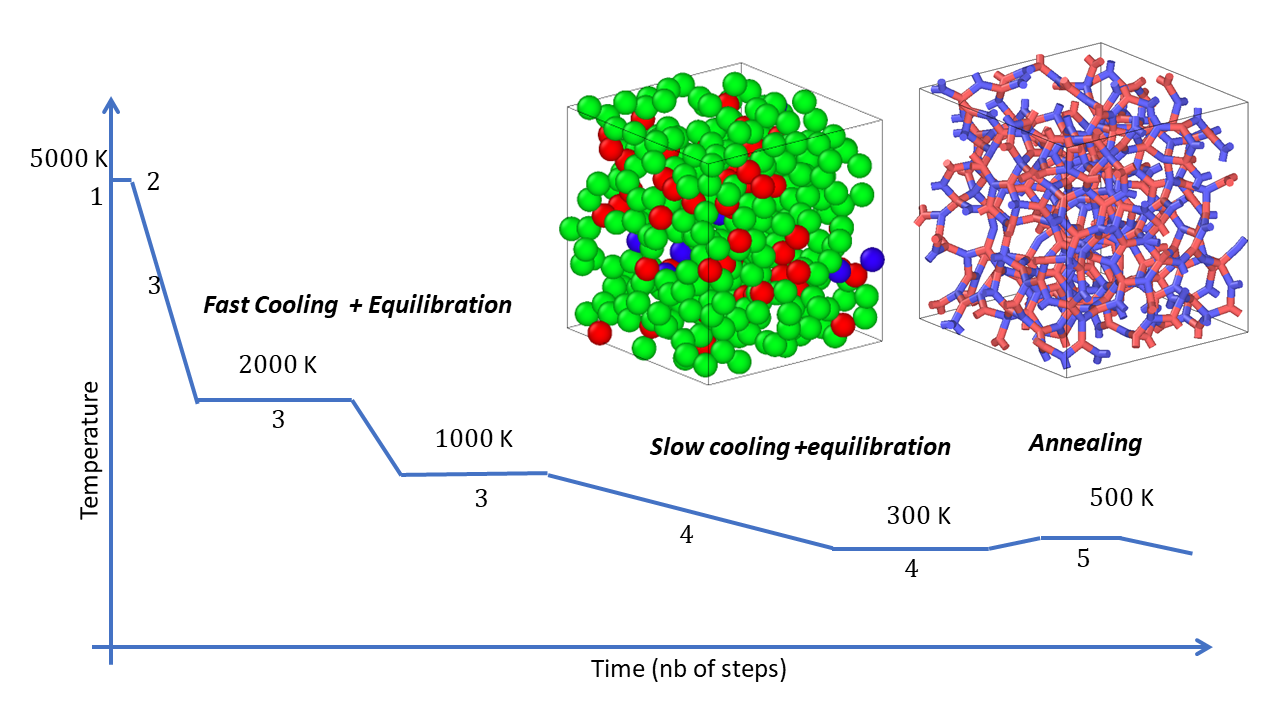}
    \caption{Melt-quenching protocol of aBN samples used GAP-MD simulations (temperatures and times are not to scale). Insets: aBN sample with 500 atoms where ${\rm sp^1, sp^2}$, and ${\rm sp^3}$-hybridized atoms are shown as blue, green, and red, respectively. Bonding of B and N atoms in an aBN sample. }
    \label{fig:MDprotocol}
\end{figure}

We employ a melt-quenching protocol to generate samples, as presented in Fig.\ref{fig:MDprotocol}, using the Large-scale Atomic/Molecular Massively Parallel Simulator (LAMMPS) and machine-learning based GAP model generated by Kaya et al. \cite{Kaya2023}. Here, we first place an equal number of B and N atoms in a simulation box (1) and melt the samples at 5000 K (2). Later, the samples are cooled down to 2000 K fast, after a short equilibration run, and then quenched to 1000 K with a high cooling rate (3). The samples are later cooled down to 300 K by employing different cooling rates (4). Finally, we employ an annealing step at 500 K to reduce the amount of $\rm sp^1$-hybridized atoms and homonuclear bonds (5).

In this section, we will focus our analysis on two different samples obtained with different cooling rates (sample aBN2 is quenched faster than sample aBN1) and annealing rates (the properties of sample aBN1 are explored for different annealing rates, from aBN1-0 having the slowest annealing to aBN1-3 the fastest). Detailed values of their cooling rates are given in table \ref{table:MDCoolingRates}.

\begin{table}[]
 \centering
 \begin{tabular}{ rrrrrr  }
  \hline

   Structure & aBN1-0 & aBN1-1 & aBN1-2 & aBN1-3 & aBN2 \\ 
   Fast cooling (K/ps) & $20$ &   $20$ & $20$ & $20$ & $50$ \\
   Slow cooling (K/ps) & $10$ &   $10$ & $10$ & $10$ & $25$ \\
   Annealing (K/ps) & $5$ &   $10$ & $20$ & $25$ & $5$ \\
  
  \hline
 \end{tabular}
 \caption{Cooling rates for the MD aBN samples studied in this section. See text and figure \ref{fig:MDprotocol} for the meaning of the different rates.}
 \label{table:MDCoolingRates}
\end{table}

Finally, we visualize the samples as in Fig.\ref{fig:MDprotocol} using the code OVITO\cite{ovito}. Samples aBN1-x, which correspond to the same initial structure at different times in the annealing process, have similar morphological properties. While sample aBN1-0 (longest annealing time) has only B-N bonds and mostly $\rm sp^2$-hybridized atoms, other aBN1-x samples have a low amount of N-N bonds ($<5\%$). The amount of $\rm sp^3$-hybridized atoms are 6.0\% for each aBN1-x sample and the ratio of $\rm sp^1$ and $\rm sp^2$-hybridized atoms are between 17.1\%-20.2\% and 73.8\%-76.9\%, respectively. While the ratio of $\rm sp^1$ is 20.2\% for sample aBN1-3, it is reduced slightly with a higher cooling rate during annealing. 
Samples aBN1-x and aBN2 all have a density of 1.967 $\rm \pm 0.01$. The composition of sample aBN2 is 7.2\% $\rm sp^1$, 84.4\% $\rm sp^2$ and 8.4\% $\rm sp^3$, and nearly all bonds are B-N bonds ($>95\%$). Even though this sample has the same density as the aBN1-x samples, there are several noticeably large microvoids within the structure. 

\subsection{The dielectric function}
\label{ss:TB_DielectricFunction}

The dielectric response of the system is described by its dielectric function, $\epsilon(\vb{q},\omega)$. In the single particle approximation and within the long wavelength limit ($\vb{q}\to\vb{0}$), the electronic contribution to the latter is given by \cite{Grosso2013solid}:
\begin{equation}
\label{eq:epsilon}
\epsilon(\vb{0},\omega) = 1+\frac{8\pi}{\Omega}\frac{\hbar^2}{{m_e}^2}\sum_{\alpha,\beta} \frac{\abs{\mel{\beta}{\vec{e}\cdot\hat{\vb{p}}}{\alpha}}^2}{\qty(E_\beta-E_\alpha)^2}\frac{f(E_\alpha)-f(E_\beta)}{E_\beta-E_\alpha-\hbar\omega -i\eta}
\end{equation}
where $\Omega$ is the system's volume, $m_e$ is the electron mass, $\vec{e}=\frac{\vb{q}}{\abs{\vb{q}}}$ is the electric field's polarization direction, $\hat{\vb{p}}$ is the momentum operator, the $(E_\alpha, \ket{\alpha})$ and $(E_\beta, \ket{\beta})$ are the system's electronic eigenpairs and $\eta\to0^+$ acts as a phenomenological broadening. Since we remain in the long wavelength limit in this work, we will omit the $\vb{q}$-dependence the dielectric function below and write simply $\epsilon(\omega)$ for $\epsilon(\vb{0},\omega)$. 

It is fruitful to consider separately the real and imaginary part of $\epsilon(\omega)$, denoted here respectively $\epsilon_1(\omega)$ and $\epsilon_2(\omega)$. $\epsilon_1(\omega)$ describes the system's dielectric screening, and the low-frequency dielectric constant, which is the figure of merit in this work, is given by $\epsilon_1(\omega=0)$. At zero temperature, and in the $\eta\to0^+$ limit, we have:
\begin{equation}
\label{eq:epsilonReal}
\epsilon_1(\omega) = 1+\frac{16\pi}{\Omega}\frac{\hbar^2}{{m_e}^2}\sum_{v,c} \frac{\abs{\mel{c}{\vec{e}\cdot\hat{\vb{p}}}{v}}^2}{\qty(E_c-E_v)^2}\frac{E_c-E_v}{\qty(E_c-E_v)^2-\qty(\hbar\omega)^2}
\end{equation}
where $v,c$ refer respectively to the valence (filled) and conduction (unfilled) states in the system. In the same conditions, the imaginary part of the dielectric function is given by:
\begin{equation}
\label{eq:epsilon_imag}
\epsilon_2(\omega) = \frac{8\pi}{\Omega}\frac{\hbar^2}{{m_e}^2}\sum_{v,c} \frac{\abs{\mel{c}{\vec{e}\cdot\hat{\vb{p}}}{v}}^2}{\qty(E_c-E_v)^2}\delta(E_c-E_v-\hbar\omega)
\end{equation}
$\epsilon_2(\omega)$ is typically related to the system's absorption spectrum. While the methods discussed in this work are not necessarily suitable to precisely access optical properties, $\epsilon_2(\omega)$ nevertheless provides insight on the contributions to $\epsilon_1(\omega=0)$. Indeed, it can be seen from equation \ref{eq:epsilon_imag} that it provides an energy-resolved description of the optical matrix elements $\frac{\abs{\mel{c}{\vec{e}\cdot\hat{\vb{p}}}{v}}^2}{\qty(E_c-E_v)^2}$ that make up $\epsilon_1(\omega=0)$.  This can also be seen from the Kramers-Krönig relation \cite{Grosso2013solid}:
\begin{equation}
\label{eq:KramersKronig}
\epsilon_1(\omega) = 1+\frac{2}{\pi}\int_0^{+\infty}\frac{\omega^\prime}{{\omega^\prime}^2-\omega^2} \epsilon_2(\omega^\prime) d\omega^\prime
\end{equation}

\subsection{A simple tight-binding model}

To access the electronic properties of aBN in a qualitative way, we rely on a simple tight-binding model. We employ a first nearest neighbor Slater-Koster model \cite{SlaterKoster1954} fitted on the cubic, wurtzite and single-layer hexagonal allotropes of BN at equilibrium and under $10\%$ isotropic dilation to determine the Slater-Koster parameters and their dependence on interatomic distances. B-B and N-N hoppings are parametrized \textit{ad hoc}, using the equivalent B-N parameters as order of magnitude estimates. Details of the fitting procedure and parameters are discussed in \ref{a:TB_SK}. We must stress that, while the tight-binding methodology accounts correctly for the different local coordination numbers and geometries, this model does not include any energetic corrections for them apart from the aforementioned distance-dependent hoppings. It is therefore limited only to a qualitative exploration of the phenomena at play and will serve to guide intuition. In fact, we found that a minimal one-orbital toy-model displayed similar trends. We briefly discuss this other model in \ref{a:TB_toy}.

\begin{figure}
 \begin{subfigure}{0.49\textwidth}
   \includegraphics[width=\textwidth]{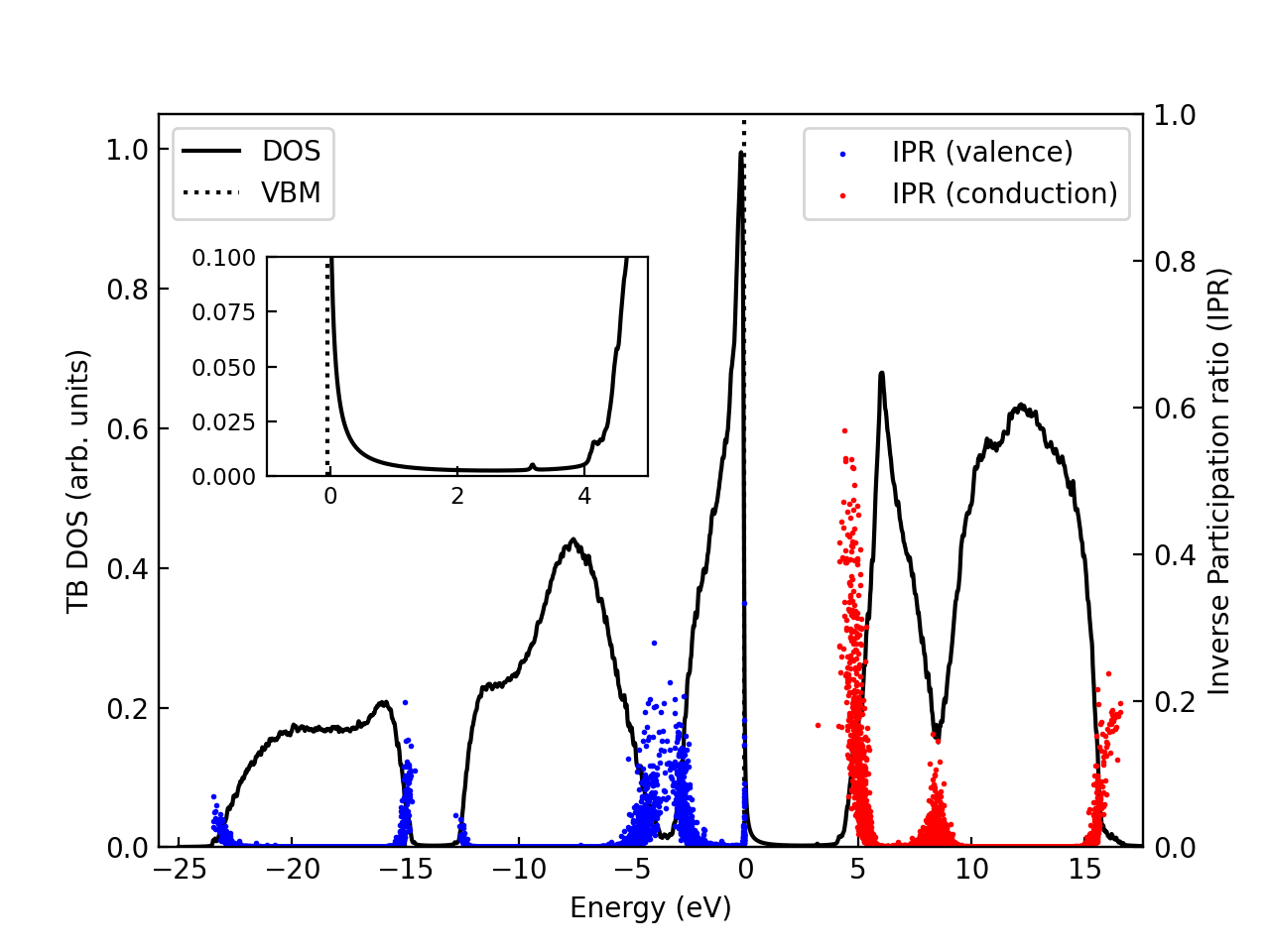}
     \caption{ }
 \end{subfigure}
 \hfill
 \begin{subfigure}{0.49\textwidth}
   \includegraphics[width=\textwidth]{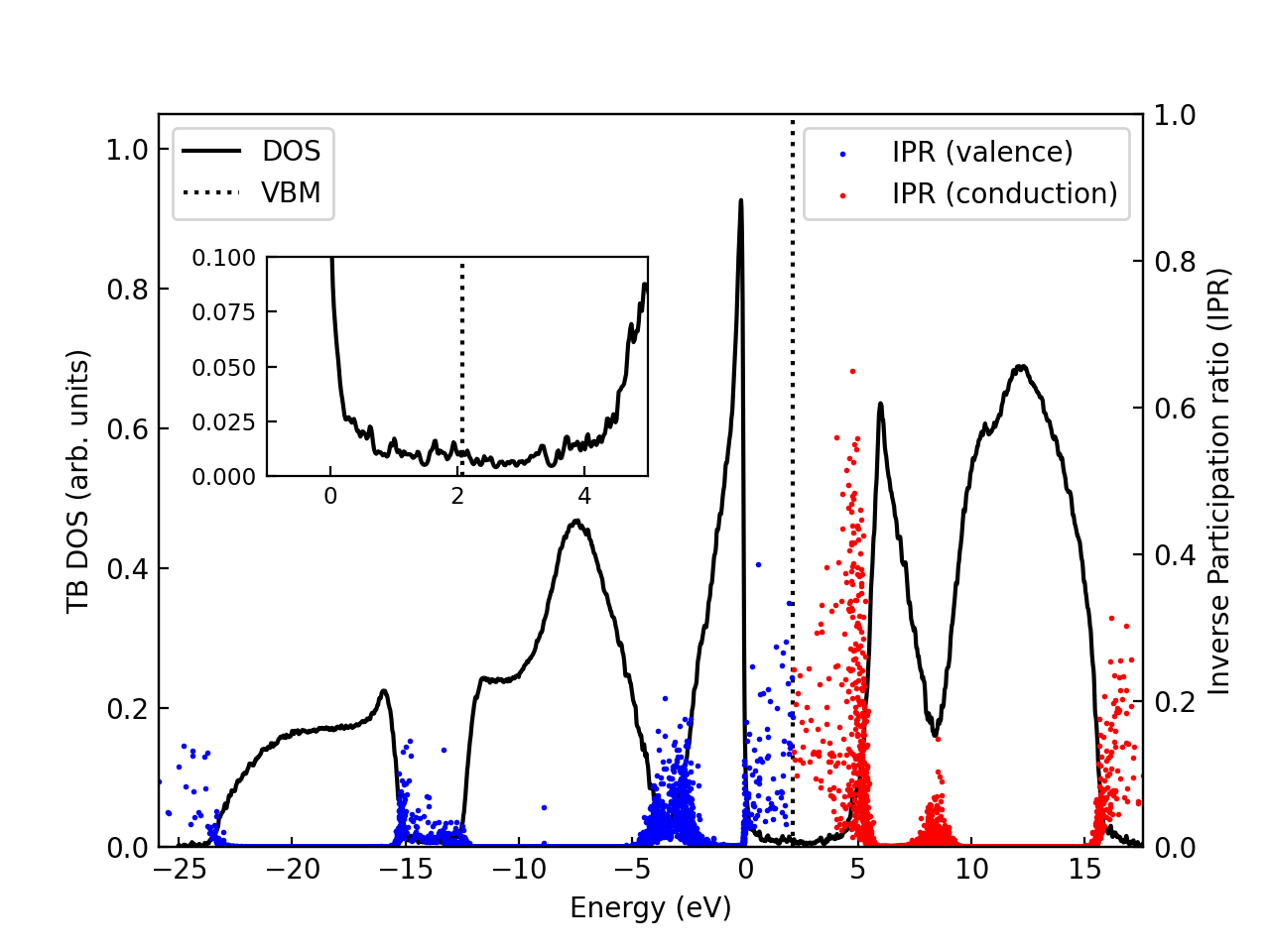}
     \caption{ }
 \end{subfigure}
 \caption{Tight-binding density of states (DOS, in black) and Inverse Participation Ratio (IPR, blue and red dots for valence and conduction states respectively) for a $\sim 10^4$ atoms aBN sample with equal concentration of Boron and Nitrogen. States are present within the electronic gap (between about $0$ and $5$ eV in this simple model), but their high IPR suggests that they are localized in real space. Panels (a) and (b) show respectively samples aBN1-0 and aBN2 (aBN2 is quenched faster than aBN1); note how a faster quenching step increases the amount of mid-gap states. Insets: zoom on the DOS of the corresponding samples in the mid-gap region. the dashed line displays the energy of the last occupied state.}
 \label{fig:DOS_IPR}
 \end{figure}

Figure \ref{fig:DOS_IPR} presents the density of states (DOS) of two $11520$ atoms structures generated using the GAP methodology (see Section \ref{s:StructuralModels}) with different quenching rates.\footnote{Calculations were performed using exact diagonalization with periodic boundary conditions, sampling the $\Gamma$ point only, with a Lorentzian broadening of the DOS of $\eta=26 \text{meV}$. See \ref{a:TB_numerics} for details.} Mid-gap states are present in both cases, although in manifestly lesser amounts than in the DFT results of Section \ref{s:SmallDFT}. 
This discrepancy may be due either to the better relaxation of these large structures, in the sense that they are less constrained by small-cell periodic boundary conditions, to the lesser statistical impact of single ``defects" in a comparatively large structure with a broad range of local environments, or by the inability of the tight-binding model to fully describe the energetic correction associated to them.
Even with these limitations, we do observe markedly more mid-gap states in the sample with the faster quenching rate (aBN2): in fact, the electronic gap effectively vanishes in this sample, as suggested by the averaging procedure for small samples performed in \ref{fig:aDOS}. In contrast, the slowly quenched sample (aBN1) retains an electronic gap of about $4-5 \text{ eV}$, reduced from its value of $6.1 \text{ eV}$ for the reference crystalline hBN in our simple model (see hBN PSL in Section \ref{ss:DielectricResponseTB}). A similar effect occurs for samples aBN1-x, for which we observe a more marked population of mid-gap states for shorter annealing times, although these samples always retain an electronic gap.

In Section \ref{s:SmallDFT}, we had suggested that the mid-gap states may be localized in real space, but the small size of the samples made it difficult to explore that hypothesis. Having now moved to larger samples, we investigate this point by computing the Inverse Participation Ratio (IPR) for all eigenstates as an indicator of localization. For a given eigenstate $\Psi$, we compute its IPR as:
\begin{equation}
\mathrm{IPR}(\Psi) = \sum_{\vb{n}} \qty(\sum_\mu\abs{\braket{\mu,\vb{n}}{\Psi}}^2)^2
\end{equation}
where the first sum is taken over atomic positions and the second over the corresponding orbitals, with $\ket{\mu,\vb{n}}$ being the orbital $\mu\in\qty{B_{2s}, B_{2p_x},B_{2p_y},B_{2p_z}}$ or $\qty{N_{2s},N_{2p_x}, N_{2p_y},N_{2p_z}}$ at position $\vb{n}$. Heuristically, if a state is delocalized over the whole system, i.e. if it has a Bloch-wave type behavior, $\abs{\braket{\mu,\vb{n}}{\Psi}}^2 \approx \frac{1}{N}$ where $N$ is the number of orbitals in the system, and therefore $\mathrm{IPR}(\Psi) \approx N\times\frac{1}{N^2}\to 0$. However, for localized states, which in the extreme case have probability densities of the form $\abs{\braket{\mu,\vb{n}}{\Psi}}^2 \approx \delta_{n,n_0}$ for some position $n_0$, the IPR is finite (and of order unity).

The results are reported in figure \ref{fig:DOS_IPR}. It can be seen in particular that, while IPR within ``bands" are almost zero, suggesting delocalized states, the IPR for states at the band-edges, and, notably, of the mid-gap states are of order unity. This suggests that these states are indeed localized in real space, and consequently should not or only weakly contribute to DC conduction in the system \cite{DRABOLD1998153}. However, their impact on the system's dielectric constant is less clear. On the one hand, as can be recovered from equation \ref{eq:epsilonReal}, low-energy transitions such as those between mid-gap states can provide a strong contribution to $\epsilon_1$, due to the presence of $(E_c-E_v)^3$ in the denominators. On the other hand, transitions between localized states are typically difficult because the optical matrix elements between them, the $\abs{\mel{c}{\vec{e}\cdot\hat{\vb{p}}}{v}}^2$, are essentially local in real space and can be thus suppressed for the same reasons that the DC conductivity is suppressed (if the valence and localization states in a given $(v,c)$ pair are localized at different positions in the solid, their overlap is small). In fact, it has been shown that, close to the Anderson transition, the dielectric constant of 3D systems should approximately correlate with the square of the wave function localization length \cite{Feigel2018}. Even in this case, however, (optical) transitions between localized mid-gap states and extended ``band-like" states are \textit{a priori} still possible, and because $\epsilon_1$ integrates transitions from all energies, the presence of mid-gap states, even localized, can increase it.

\subsection{Dielectric response}
\label{ss:DielectricResponseTB}

To investigate these questions more directly, we now turn to the computation of $\epsilon(\omega)$. To this end, we use a common tight-binding approximation to express the momentum matrix elements in terms of the Hamiltonian matrix elements \cite{Delerue2013nanostructures}:
\begin{equation}
\label{eq:momentumTB}
\mel{\mu,\vb{n}}{\hat{\vb{p}}}{\mu^\prime,\vb{n^\prime}} = -\frac{m_e}{i\hbar}\qty(\vb{n^\prime}-\vb{n})\mel{\mu,\vb{n}}{\hat{H}}{\mu^\prime,\vb{n^\prime}}
\end{equation}
Equation \ref{eq:momentumTB} also substantiates our earlier remark about the local character of the momentum matrix elements, which in this tight-binding formulation can be seen to be a direct consequence of the local character of the tight-binding Hamiltonian (itself a consequence of the exponential decay of the atomic orbitals).

\begin{figure}[ht]
    \centering
    \includegraphics[width=0.9\columnwidth]{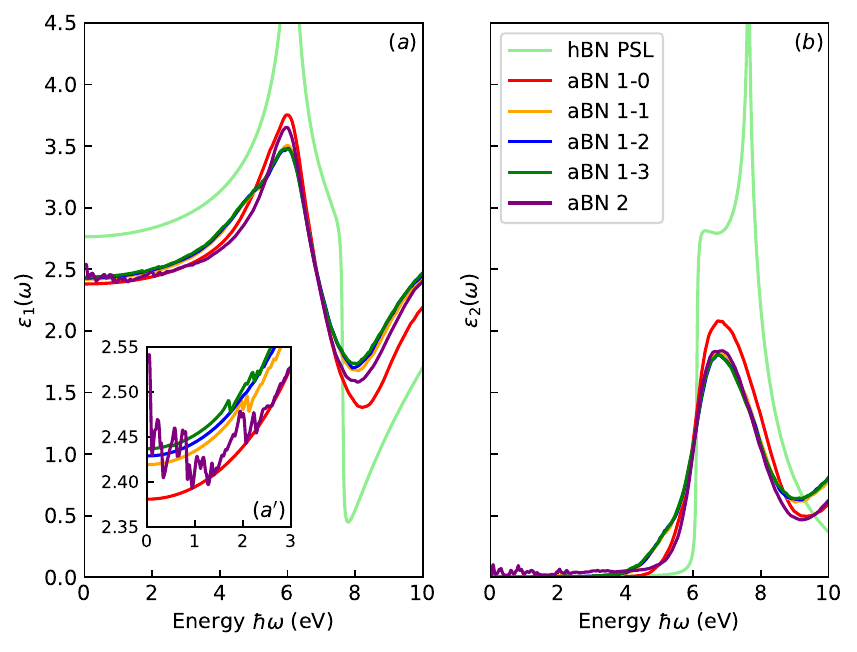}
    \caption{Real (panel (a)) and imaginary (panel (b)) parts of the dielectric function for the aBN sample generated with different quenching and annealing time (see text and DOS displayed in figure \ref{fig:DOS_IPR}). The light green curve (hBN PSL) corresponds to the response of periodically stacked hBN single layer (see text). Inset (a') displays a zoom on the low-frequency part of $\epsilon_1$, showing how a longer annealing or a shorter quenching time may lead to a lower dielectric constant. All panels share their legend. The higher energy behavior of $\epsilon_2\qty(\omega)$ and a more detailed analysis of transitions are given in \ref{a:KramersKronig}.}
    \label{fig:Epsilon_SK}
\end{figure}

Figure \ref{fig:Epsilon_SK}  displays the dielectric function for samples aBN1-x and aBN2 (the DOS of aBN1-0 and aBN2 are also shown in figure \ref{fig:DOS_IPR}).\footnote{See \ref{a:TB_numerics} for computational details.} As expected from the tendency of large systems to self-average, we find that at these scales ($\sim 10^4$ atoms) the dielectric properties of the system are essentially isotropic.\footnote{It should be noted that our sample generation procedure does not include the presence of a substrate. In real systems, and for thin enough films, substrate interaction could induce an anisotropic response, e.g. by favoring alignment of BN rings.} We therefore represent the average of $\epsilon$ over the cartesian directions. To provide a basis for comparison, we also display the dielectric function for bulk hBN averaged in such a manner, using the same model. Because the hoppings in our tight-binding model have a cutoff radius that is inferior to the interlayer distance in hBN, interlayer hoppings are not taken into account for consistency. We thus call this reference system ``Periodically-stacked single layer hBN" (hBN PSL) for clarity. It also results from this that the dielectric constant of hBN PSL reduces to $1$ along the stacking direction.\footnote{As a consequence, the averaged dielectric function which we display for hBN PSL for consistency is given by $\bar{\epsilon}=1+\frac{2}{3}\qty(\epsilon_\parallel-1)$, where $\epsilon_\parallel$ is its dielectric constant for in-plane polarized fields.} In effect, this ``stacking" thus only enters the computation of $\epsilon$ through the interlayer distance, which affects the system's volume $\Omega$ (see equation \ref{eq:epsilon}).

Panel (a) displays $\epsilon_1(\omega)$. The figure of merit, here, is the static dielectric constant $\epsilon_1(\omega=0)$. We must stress that in the context of aBN, the low-frequency dielectric constant in capacitance experiments is typically measured at frequencies of the order of hundreds of kHZ\cite{Hong2020,Lin2022}, which correspond to energies $\hbar\omega$ of the order of $10^{-9} \text{ eV}$. Variations of $\epsilon_1(\omega)$ on these energy scales likely have origins in vibrational contributions to $\epsilon_1$, which are not included in our frozen-atoms calculations. With this caveat, it can be seen that, in this model, aBN does have a lower average dielectric constant than our reference crystalline system. We also note that $\epsilon_1$ decreases as the sample's annealing or quenching time is increased, as can be seen in the inset (a'), although the effect in this formulation is relatively weak.

To understand these effects we turn to the imaginary part of the dielectric function, which is displayed in panel (b). As per equation \ref{eq:epsilon_imag}, $\epsilon_2(\omega)$ resolves in energy both the density of transitions and their strength (it is in effect the joint density of states weighted by oscillator strengths). Optically active transitions can be observed below the transition edge of the crystalline reference system, which by definition can be ascribed to the contribution of mid-gap states. We also observe the suppression of the van-Hove singularities, which is symptomatic of the loss of crystal order in the samples. In addition, and in contrast to what happens below the crystalline gap, there is a sizeable reduction of the values of $\epsilon_2\qty(\omega)$ above the reference crystal gap, which, through equation \ref{eq:epsilon_imag}, can be interpreted as an overall weakening of oscillator strengths in the system. Because the overall oscillator strength-density of mid-gap states is low in these models, even if they have lower transition energies, the net effect through equation \ref{eq:epsilonReal} or \ref{eq:KramersKronig} is a reduction of the static dielectric constant compared to the reference crystalline sample. This is further shown through the whole energy range of transitions by a Kramers-Kr\"onig analysis in \ref{a:KramersKronig}.

\subsection{Beyond current modeling}

The above results suggest that there is a competition between two effects. On one side, as was discussed in section \ref{ss:SmallSamplesConclusion}, disorder in the samples has a tendency to increase the dielectric constant, likely through the emergence of mid-gap states. This effect is also observed in this section, although only weakly in our simple TB model. On the other side,  as seen above, the amorphous structure of the sample yields an overall reduction in the oscillator strength of its transitions compared to our crystalline reference, which results in a lowering of its dielectric constant. In addition, disorder induces localization effects which are visible in large samples, and should contribute to a lowering of the optical matrix elements. To obtain a clearer understanding of these combined effects, it thus seems necessary to possess a good description of the mid-gap states, and more generally of the energetics of the numerous atomic configurations and local environments present in the samples (onsites, hoppings...). This description is \textit{a priori} available to \textit{ab initio} techniques, but it is lacking in our simple tight-binding model. Conversely, the large-scale effects of disorder that we just discussed are very computationally demanding for \textit{ab initio} methods. It thus appears desirable to move towards more sophisticated tight-binding (or generally second principles) models. Finally, it is believed that amorphous BN materials also contain hydrogen or oxygen atoms in non negligible quantity, impacting their structural stability and mechanical properties \cite{kaya2023impact}, so that one should also consider proper tight-binding modeling of these additional atomic species for accessing the full dielectric response of measured samples.

For completeness, we make here a remark and a further caveat. Owing to the complexity and size of the amorphous systems considered here, we have remained at the theoretical level of (effectively) independent particles. It is well known, however that in hBN, quasiparticle corrections and electron-hole interaction effects (excitons) are far from negligible \cite{Blase1995, Arnaud2006}. The effect of such many-body corrections is typically larger if the Coulomb potential is poorly screened, making aBN a likely candidate to display such effects despite its 3D nature. While even in such systems, DFT methods have been used to compute quantities such as the static dielectric constant \cite{Laturia2018}, they are typically insufficient to describe their optical properties. The same limitation applies to single-particle tight-binding methods such as the ones used here. Nevertheless, as per Section \ref{ss:TB_DielectricFunction}, we believe that the $\omega>0$ behavior of the dielectric function, even at this level of theory, can be helpful both as a first step in this direction, and in understanding the static trends. Obtaining an accurate estimation of $\epsilon(\omega>0)$ which could be compared with the results of optical experiments is however likely to necessitate some form of the aforementioned many-body corrections. This is however out of the scope of this work.

To conclude this section, we point out that significant work remains to be done to reach accurate modeling of aBN systems. In this context, it is not so surprising that the values of the static dielectric constants computed in this work remain above two, while experimental values of $\epsilon_1 \approx 2$ and below have been reported in the litterature \cite{Hong2020,Lin2022}.

\section{Conclusions and perspectives}

We report on some exploratory modeling studies towards the understanding of electronic and dielectric properties in models of amorphous Boron Nitride. We combined the analysis of a dataset of small ($\sim 10^2$ atoms) samples at the DFT level with the study of large ($\sim 10^4$ atoms) samples generated with machine learning interatomic potentials whose electronic properties have been explored using a simple tight-binding model. In both cases, we have observed the formation of mid-gap states inside the otherwise large ionic gap (as observed in clean hBN for example).
The density of these states, and the dielectric constant of the samples are tentatively related to the system's atomic density, and to the nature of its atomic bonding characteristics. For small samples, we highlight in particular that, in ${\rm sp^2}$ dominated systems, the presence of ${\rm sp^1}$ and ${\rm sp^3}$ coordinated atoms appears to act as a source of extra disorder which increases the dielectric constant. Likewise, departure from a B-N alternating structure or a B/N=1 stoichiometric ratio are also observed to have detrimental effects on the system's dielectric constant.
However, the disordered nature of these materials demands to consider very large scale effects, which are out of scope for first-principles simulations.
We have thus analyzed large systems through the prism of a simple tight-binding Hamiltonian, and, in this model, found them to display an overall weakening of their transitions's oscillator strengths due to their amorphous nature, which in turn lowers their dielectric constant. We also observe {\it real space localization effects} for the mid-gap states, which may lead to a partial suppression of their contribution in low-energy transitions and therefore a lower contribution to dielectric properties.
To obtain a more realistic and complete view of the competing phenomena at play, efforts are therefore further needed to develop more refined TB Hamiltonian, retaining the physics related to the bonding effects in the various observed structural conformations. It will demand adequate adjustment of parameters with {\it ab-initio} data on smaller system sizes, eventually harnessing the power of machine learning techniques to automatize the extraction of tight-binding parameters in an arbitrary complex disordered structure. Such models seem particularly appropriate in the context of aBN, as tight-binding naturally accounts for the geometric structure of the system, while benefiting from well-established linear scaling methods \cite{PhysRevLett.79.2518,PhysRevB.59.2284,Fan2021}, which allow the study of extremely large samples ($>10^6$ atoms).
The use of Artificial Intelligence-based techniques, meanwhile, has become key to obtain realistic electronic models able to deliver quantitative predictions \cite{Schleder_2019,Li_2022}.
This will be fundamental to reach a predictive modeling of dielectric constants or other properties in amorphous forms of boron-nitride composites, enabling further materials optimization and performance improvement \cite{Matsoso_2021,Molina-Garcia_2023,Kaya2024b}.

\section*{Acknowledgements}

This project has been supported by Samsung Advanced Institute of Technology. This paper reflects only the author's view and the Research Executive Agency is not responsible for any use that may be made of the information it contains. ICN2 acknowledges the Grant PCI2021-122092-2A funded by MCIN/AEI/10.13039/501100011033 and by the “European Union NextGenerationEU/PRTR”. This work has been performed “Amb el suport del Departament de Recerca i Universitats de la Generalitat de Catalunya” and with funding from PN: “Proyecto PID2022-138283NB-I00 (MCIN/ AEI /10.13039/501100011033/) and by FEDER. The MINERVA project is also aknowledged: “Proyecto PCI2021-122092-2ª financiado por MCIN/AEI /10.13039/501100011033 y por la Unión Europea NextGenerationEU/PRTR”. This project is also funded by the European Union's Horizon 2020 Research and Innovation programme under the Marie Sklodowska-Curie entitled STiBNite (N$^{\circ}$ 956923), by the F\'ed\'eration Wallonie-Bruxelles through the ARC on Dynamically Reconfigurable Moir\'e Materials (N$^{\circ}$ 21/26-116), by the Flag-Era JTC project “MINERVA” (N$^{\circ}$ R.8006.21), by the EOS project “CONNECT” (N$^{\circ}$ 40007563) and by the Belgium F.R.S.-FNRS through the research project (N$^{\circ}$ T.029.22F).
Simulations were performed at the Center for Nanoscale Materials, a U.S. Department of Energy Office of Science User Facility, supported by the U.S. DOE, Office of Basic Energy Sciences, under Contract No. DE-AC02-06CH11357.
Computational resources have also been provided by the CISM supercomputing facilities of UCLouvain and the C\'ECI consortium funded by F.R.S.-FNRS of Belgium (N$^{\circ}$ 2.5020.11). 
Additional computational support was received from the King Abdullah University of Science and Technology-KAUST (Supercomputer Shaheen II Cray XC40) and Texas Advanced Computing Center (TACC) at The University of Texas at Austin. O.K. is supported by the REDI Program, a project that has received funding from the European Union's Horizon 2020 research and innovation program under the Marie Skłodowska-Curie grant agreement no. 101034328.

\appendix

\section{Slater-Koster tight-binding model}
\label{a:TB_SK}

In this study, a simple tight binding model is used to gain qualitative insights into the electronic and dielectric response of amorphous Boron Nitride. The model relies on  atomic orbitals from $s$ and $p$ subshells (i.e. 4 orbitals per atomic site) to expand the low energy electronic manifold. An effective onsite energy $h^{I}_{\theta}$, with $\theta \in \{ s, p\}$ and $I \in \{B, N\}$,  is associated with each atomic subshell so as to parameterize the diagonal Hamiltonian matrix elements. The non-diagonal Hamiltonian matrix elements are described as two-center integrals and parameterized according to the Slater-Koster formalism,   
 \begin{equation}
t_{\gamma} (r) =
\begin{cases}
\ t_{\gamma} e^{-\beta_{\gamma} \left(\frac{r}{r_0} - 1 \right)} &\mbox{if $r<r_c$}  \\
\ 0 &\mbox{otherwise}
\end{cases}
\end{equation}
where $t_{\gamma} \in \{t_{ss_{\sigma}}, t_{sp_{\sigma}}, t_{ps_{\sigma}}, t_{pp_{\sigma}}, t_{pp_{\pi}}\}$ are the Slater-Koster parameters associated with electron hoping between Boron and Nitrogen atoms, $r$ is the atomic distance ($r_0 = 1.57 \AA$), and $\beta_{\gamma} \in \{ \beta_{ss_{\sigma}}, \beta_{sp_{\sigma}}, \beta_{ps_{\sigma}}, \beta_{pp_{\sigma}}, \beta_{pp_{\pi}}$\} are dimensionless parameters introducing functional dependence of energy integrals with respect to radial distance. The cutoff radius ($r_c$ = 1.7 \AA) restricts effectively the direct interaction to nearest-neighbour atoms. Note that $t_{sp_{\sigma}} \neq t_{ps_{\sigma}}$ as the left hand side is associated with hoping between $B_s - N_p$ orbitals, while the right hand side relates to $B_p - N_s$ hoping.

\begin{figure}
 \includegraphics[width=\textwidth]{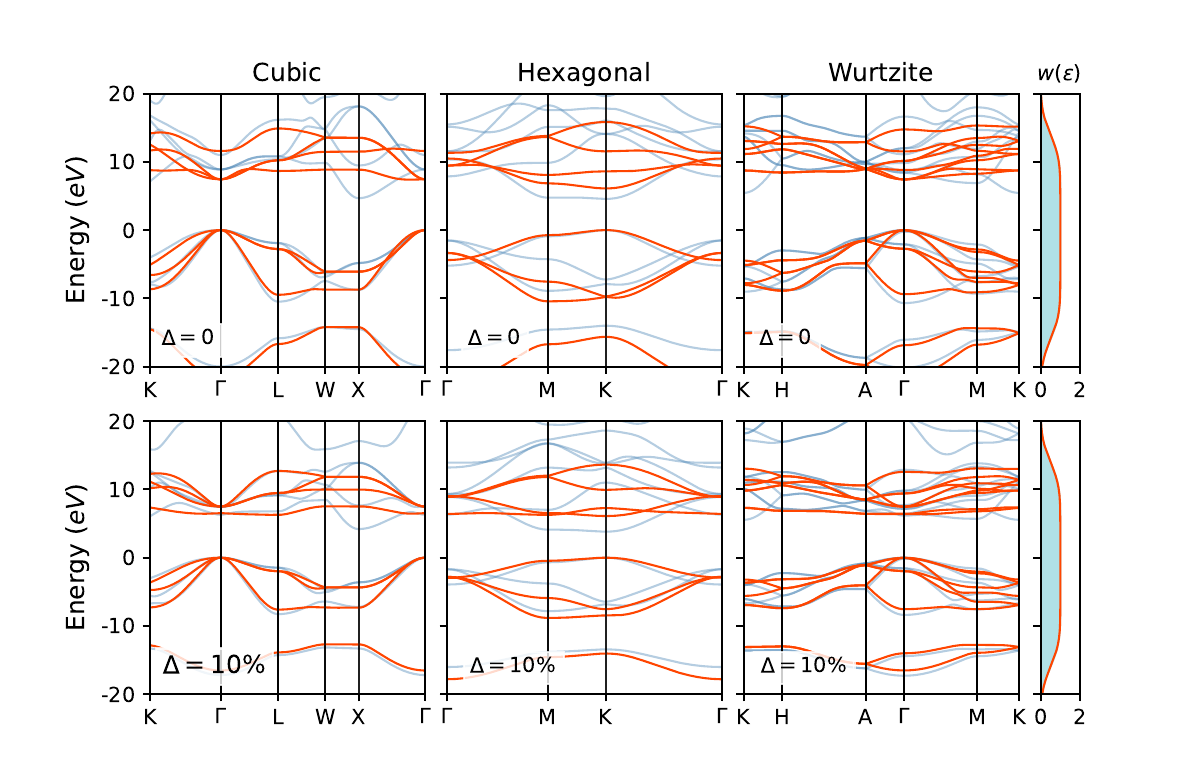}
 \caption{Band structure of allotropes of Boron Nitride used to fit the Slater-Koster parameters. Cubic, monolayer hexagonal and wurtzite allotropes (from left to right) have been considered both at equilibrium volume (top row) and under an isotropic dilation of $\Delta = 10\%$ (bottom row). The reference DFT and TB calculations  are depicted in blue and orange respectively. The weight function ($w(\epsilon)$) used to restrict the energy range of the fit is depicted on  the right.}
 \label{fig:TB_bands}
\end{figure}

The model parameters have been fitted with respect to reference electronic structure calculations by minimizing the following loss function, 
\begin{equation}
Loss  = \sum_{\Gamma} L_{\Gamma} \ \ \mbox{ with } 
L_{\Gamma} = \left[ \int_{BZ} \sum_{m} \,  w(\epsilon^{ref}_{m\mathbf{k}})\, (\epsilon^{tb}_{m\mathbf{k}} - \epsilon^{ref}_{m\mathbf{k}})^2 \   d\mathbf{k} \right]_{\Gamma} \ ,
\end{equation}
built from the difference between the reference and tight-binding eigenvalues $(\epsilon^{tb}_{m\mathbf{k}} - \epsilon^{ref}_{m\mathbf{k}})$ across the first Brillouin zone ($\mathbf{k}\in BZ$). Here, $m$ is the band index and the sum runs over all bands. The weight function $w(\epsilon)$ is introduced to restrict the energy range. The weight function used here is depicted in Fig.~\ref{fig:TB_bands}. The integral is evaluated for different atomic configurations, here labelled $\Gamma$, that contribute equally to the loss function. The set of atomic configurations considered in this work is composed of various allotropes of Boron Nitride, namely the cubic, wurtzite and monolayer hexagonal allotropes, both at equilibrium volume and under an isotropic dilation ($\Delta = 10\%$). The reference data have been computed within the framework of DFT. The equilibrium structural parameters were obtained upon geometry optimization with VASP with the same functional and cutoff energy as in Sec.~\ref{s:SmallDFT} and regular $k$-points grids of $6\times6\times6$, $6\times6\times6$ and $16\times16\times1$, respectively for the cubic, wurtzite and monolayer hexagonal allotropes. The set of reference eigenvalues used for the fitting were then obtained by self-consistently expanding the electronic manifold of the equilibrium and dilated allotropes upon a basis sets of numerical atomic orbitals (double-$\zeta$ + polarization) as implemented in the Siesta simulation package~\cite{Soler2002}. Integration in reciprocal space was performed on regular grids characterized by an effective cutoff $\geq 10 \text{ \AA}$. The plane wave cutoff for the real-space grid was set to 800 Ry. The parameters obtained after fitting are reported in Table~\ref{table:TB_params}. The band-structures computed with our tight-binding model are depicted in Fig.~\ref{fig:TB_bands}. Comparison with the reference data shows a surprisingly good accordance for the low energy valence eigenstates indicating a high degree of transferability for this manifold. The agreement is less pronounced for the conduction eigenstates. While some characteristic energy dispersion of the low energy conduction bands are well reproduced, other features associated with longer ranged interaction or with significant contribution from atomic subshells of higher angular momenta are missing. Nevertheless, our tight-binding model, despite its simple form, captures qualitatively well the energy dispersion range of the first few valence and conduction bands. 

\begin{table}[h!]
 \centering
 \begin{tabular}{ rrrrr  }
  \hline
  \multicolumn{5}{l}{Onsite parameters $[eV]$} \\
  $h^{B}_{s}$ & $h^{B}_{p}$ & $h^{N}_{s}$ & $h^{N}_{p}$ & \\ 
   -1.2 &   6.1 & -10.2 & 0. & \\
  \hline
  \multicolumn{5}{l}{Hopping parameters $[eV]$} \\
  $t_{ss_{\sigma}}$ & $t_{sp_{\sigma}}$ & $t_{ps_{\sigma}}$ & $t_{pp_{\sigma}}$ & $t_{pp_{\pi}}$ \\
  -4.0 &  3.8 & 4.3 & 5.2 & -1.8 \\
  \hline
  \multicolumn{5}{l}{Distance parameters} \\
  $\beta_{ss_{\sigma}}$ & $\beta_{sp_{\sigma}}$ & $\beta_{ps_{\sigma}}$ & $\beta_{pp_{\sigma}}$ & $\beta_{pp_{\pi}}$ \\
  4.1 & 1.7 & 2.3 & 1.9 & 3.0 \\
  
  \hline
 \end{tabular}
 \caption{Tight-binding parameters from crystalline Boron Nitride allotropes fitting.}
 \label{table:TB_params}
\end{table}

In the allotropes of BN considered above, only B-N bonds are present. This is not the case in aBN, where B-B and N-N bonds are also possible and have been observed both in simulation and experimentally \cite{Zedlitz1996}.\footnote{To our knowledge, while B-B bonds are observed experimentally, N-N bonds typically are not, although they are seen in simulations.} While we observe that, in large samples at least, such bonds are typically not dominant in number ($<5\%$ of nearest neighbor bonds for the samples of Section \ref{s:StructuralModels}), we still require a prescription to incorporate them in our tight-binding framework. Since our fit to the crystalline allotropes does not provide this information, we employ an \textit{ad hoc} parametrization with the same functional forms. For hoppings between same subshell (i.e. $s$-$s$ and $p$-$p$) we use the same hopping and distance parameters for the B-B and N-N bonds as for the corresponding B-N bonds. For the $s$-$p$ hopping terms, we use the arithmetic average of the hopping parameters $s$-$p$ and $p$-$s$ derived for BN with distance parameters set to the Harrison scaling value of $2$ \cite{Harrison1989}.

\section{One-orbital tight-binding toy model}
\label{a:TB_toy}

In this appendix, we briefly discuss a very simple tight-binding toy model which nevertheless captures many of the features discussed in Section \ref{s:TightBinding}. We consider, generally, a binary alloy with two atoms of very different electronegativities, with one effective isotropic orbital per atom. For concreteness, let the two species under consideration be $B$ and $N$, so that we work with a basis of localized orbitals $\qty{\ket{B,\vb{n}},\ket{N,\vb{m}}}_{\vb{n},\vb{m}}$ where $\vb{n}$ and $\vb{m}$ run over the Nitrogen and Boron atomic positions respectively. The effective Hamiltonian is then given by:
\begin{equation}
\hat{H}=h^B\sum_{\vb{n}}\dyad{B,\vb{n}}+h^N\sum_{\vb{m}}\dyad{N,\vb{m}}+t\sum_{\langle \vb{n},\vb{n^\prime} \rangle} g(\abs{\vb{n^\prime}-\vb{n}})\dyad{\mu,\vb{n}}{\mu^\prime,\vb{n^\prime}}
\end{equation}
In other words:
\begin{align*}
\mel{B,\vb{n}}{\hat{H}}{B,\vb{n}} &=h^B \\
\mel{N,\vb{m}}{\hat{H}}{N,\vb{m}} &=h^N \\
\mel{\mu,\vb{n}}{\hat{H}}{\mu^\prime,\vb{n^\prime}} &=tg(\abs{\vb{n^\prime}-\vb{n}}) \quad \mbox{if $\vb{n}\neq\vb{n^\prime}$}
\end{align*}
Here $h^B$ and $h^N$ are fixed onsite energies for all atoms of the given species whose difference $h^B-h^N>0$ accounts for the difference in electronegativity between B and N, $t$ is the hopping element for a reference interatomic distance which corresponds to a typical equilibrium bond length $r_0$, and $g(\abs{\vb{n^\prime}-\vb{n}})$ is a decreasing function which describes the evolution of the hopping strength with distance. In keeping with the simplicity of the model, we choose here an exponential decay with a first nearest neighbors cutoff $r_c$:
\begin{equation}
g(r)=
\begin{cases}
\exp\qty(-\beta\qty(\frac{r}{r_0}-1)) &\mbox{if $r<r_c$}  \\
0 &\mbox{otherwise}
\end{cases}
\end{equation}
For consistency, we choose values for the parameters which correspond to the $\pi$ orbitals of the Slater-Koster model used in the main text (see appendix \ref{a:TB_SK}), viz.: $h^B=h_p^B=6.1 \text{ eV}$, $h^N=h_p^N=0 \text{ eV}$, $t=t_{pp_{\pi}}=-1.8 \text{ eV}$ and $\beta=\beta_{pp_{\pi}}=3.0$ with an equilibrium bond length of $r_0 = 1.57 \text{ \AA}$ and a cutoff radius $r_c=1.9 \text{ \AA}$. The crucial parameter here is the ratio between the typical hoppings and the difference in onsite energies, $\abs{\frac{t}{h^B-h^N}}$. We have selected here specifically the parameters of $p$ orbitals, as they are typically the ones involved near the band gap for BN materials \cite{Topsakal2009}, and the $pp_{\pi}$ hopping parameters because such bonds make up the $\pi/\pi^*$ bands of hBN and we mostly consider $\rm sp^2$-dominated samples.

\begin{figure}  
 \begin{subfigure}{0.49\textwidth}
   \includegraphics[width=\textwidth]{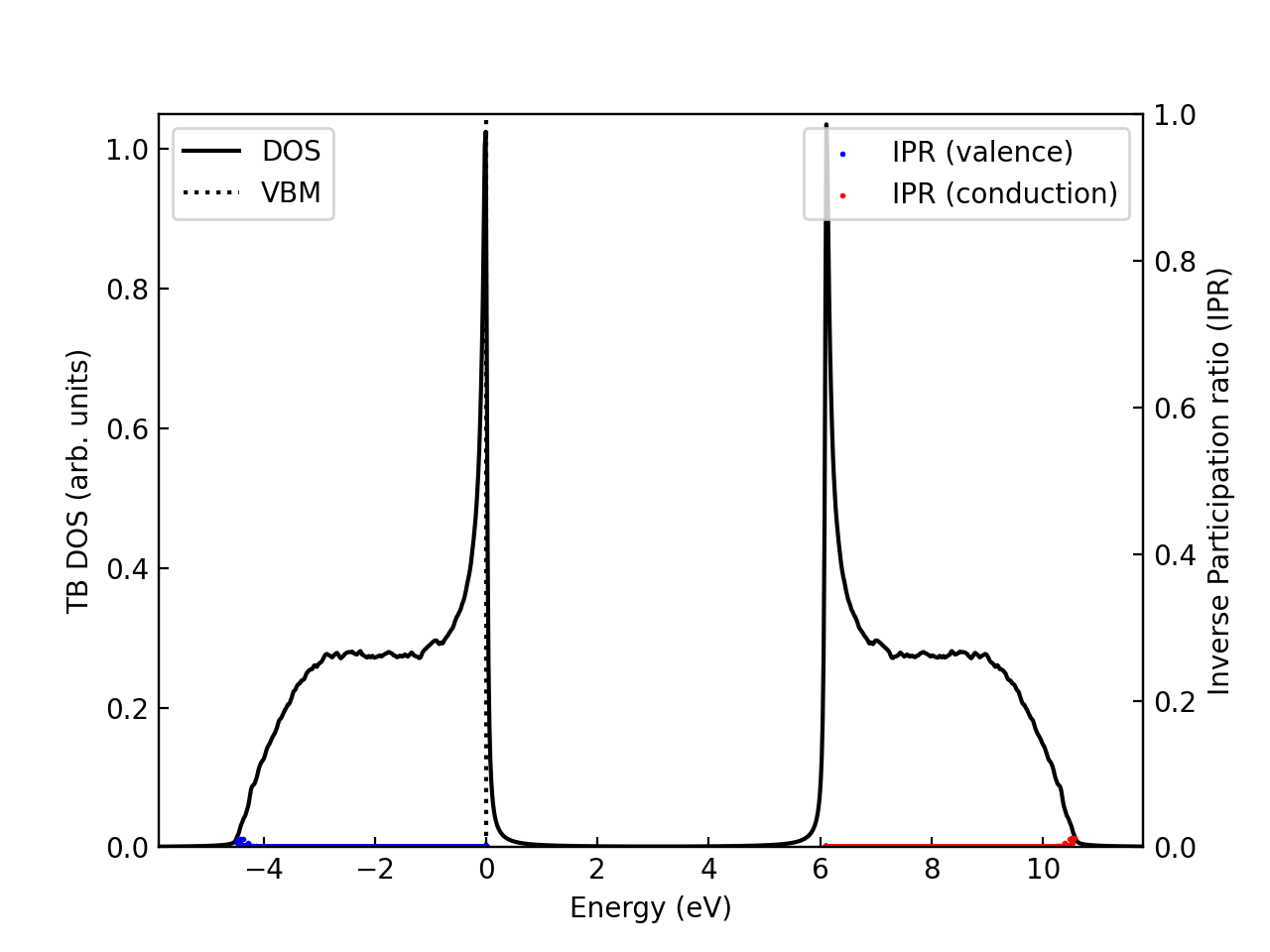}
     \caption{ }
 \end{subfigure}
 \hfill
 \begin{subfigure}{0.49\textwidth}
   \includegraphics[width=\textwidth]{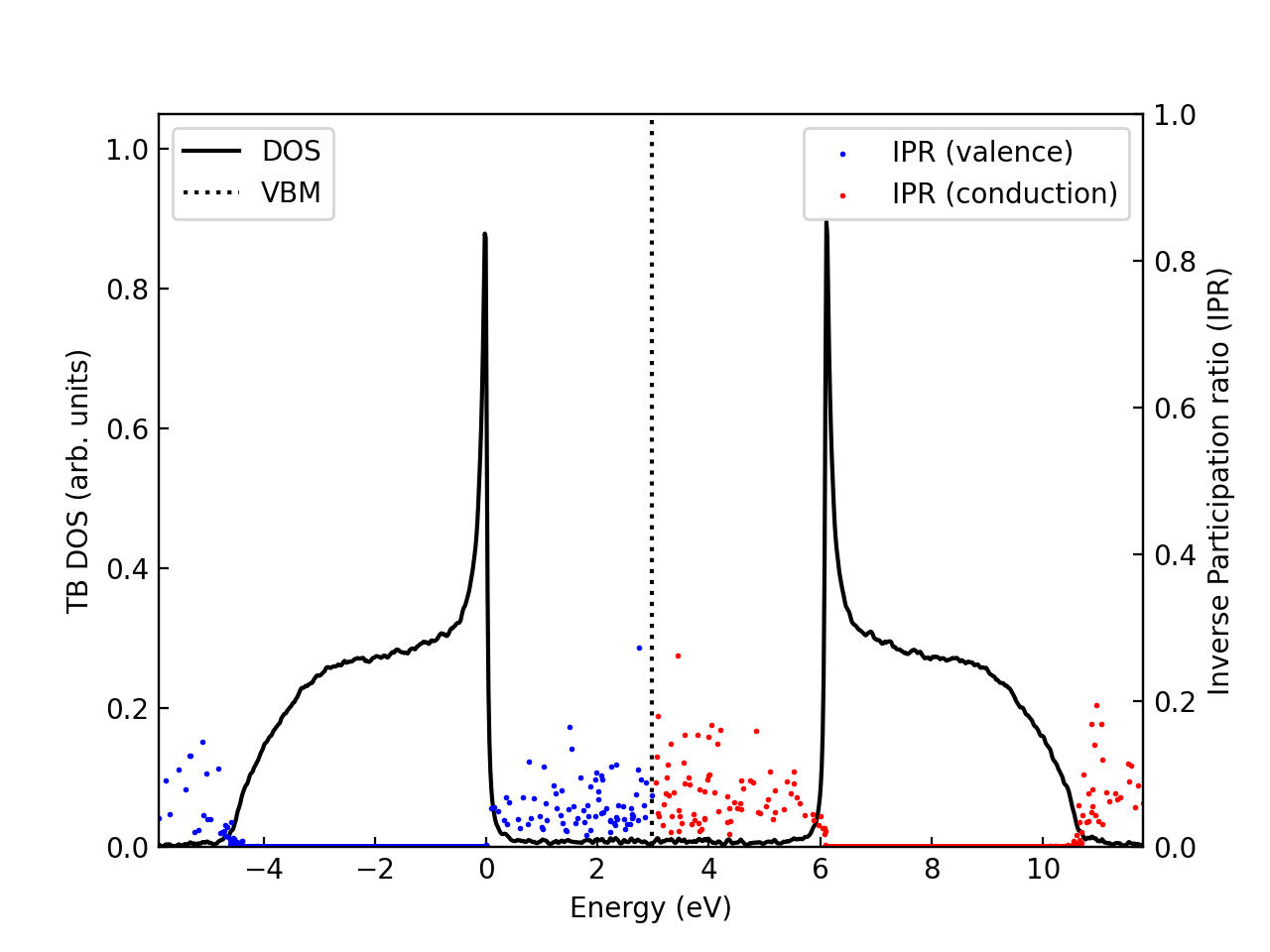}
     \caption{ }
 \end{subfigure}
 \caption{Single orbital tight-binding density of states (DOS, in black) and Inverse Participation Ratio (IPR, blue and red dots for valence and conduction states respectively) for samples aBN1-0 and aBN2 (aBN2 is cooled faster than aBN1-0) in the one orbital tight-binding toy model. Mid-gap states are observed again, and are suppressed by slower cooling rates. The dashed line denotes the energy of the highest occupied state. Figure \ref{fig:DOS_IPR} presents the DOS of the same samples using the full Slater-Koster model.}
 \label{fig:DOS_IPR-toy}
 \end{figure}

Figure \ref{fig:DOS_IPR-toy} presents the DOS and IPR in this simple model for sample aBN1-0 and aBN2 (see figure \ref{fig:DOS_IPR} displays for their DOS in the full Slater-Koster model). Except for the model, we used the same calculation parameters as in the main text (see \ref{a:TB_numerics}).
Again, localized mid-gap states are observed, although they are here completely suppressed for the slower cooled sample. We also notice that the valence and conduction DOS are essentially symmetric, except for small difference in the mid-gap states. This is at variance with the situation in the full Slater-Koster model, as seen in figure \ref{fig:DOS_IPR}. In particular, we notice there a markedly higher amount of mid-gap states on the conduction side as opposed to the valence side, while this is not the case in the one-orbital model. We hypothesize that this may be due to the relatively flat conduction bands produced by the Slater-Koster model (see figure \ref{fig:TB_bands}), whose states can thus be localized more easily under the action of disorder.

In this simple model, the presence of mid-gap states appears strongly connected to the presence of B-B and N-N bonds. In many samples, we observed a strong diminution or even a suppression of mid-gap states when artificially setting the B-B and N-N hopping integrals to zero. While this would corroborate the idea from the main text that Boron (or Nitrogen) clusters have an adverse effect on the dielectric constant, results were inconclusive when we repeated similar experiments using the full Slater-Koster model.

\section{Numerical parameters for the tight-binding calculations on aBN and hBN PSL}
\label{a:TB_numerics}

The tight-binding calculations on aBN samples and hBN PSL in section \ref{s:TightBinding} were performed by exact diagonalization of the tight-binding Hamiltonian of \ref{a:TB_SK} using periodic boundary conditions. For the $\sim 10^4$ atoms aBN samples, we sampled the Brillouin zone at the $\Gamma$ point only.

For the calculation of the DOS, a Lorentzian broadening of $26 \text{ meV}$ was used, corresponding to $k_B T$ at room temperature. We used the same phenomenological broadening of $\eta=26 \text{ meV}$ for the calculation of the dielectric function, using equations \ref{eq:epsilon} and \ref{eq:momentumTB} to compute $\epsilon\qty(\omega)$ from the eigenstates and eigenenergies obtained by diagonalization.

In the case of hBN PSL, we used experimental values for the geometry, taking $a=2.5 \text{ \AA}$ for the in-plane lattice constant (closest in-plane B-B distance) and $h=3.25 \text{ \AA}$ for the interlayer distance \cite{Fossard2017}. Because the interlayer distance is larger than the cutoff radius for the hoppings in our model, we simulated only one layer, using for convenience a rectangular 4-atoms unit cell, and we sampled the corresponding Brillouin zone with a $501\times 501 \times 1$ k-grid.

We mention that, while we used an in-house code for these calculations, we employed the Slater-Koster functions implemented in the TBPlaS code \cite{Li2023tbplas} to compute the angular dependence of the hoppings.

\section{Kramers-Kr\"onig analysis}
\label{a:KramersKronig}

\begin{figure}[ht]
    \centering
    \includegraphics[width=1.0\columnwidth]{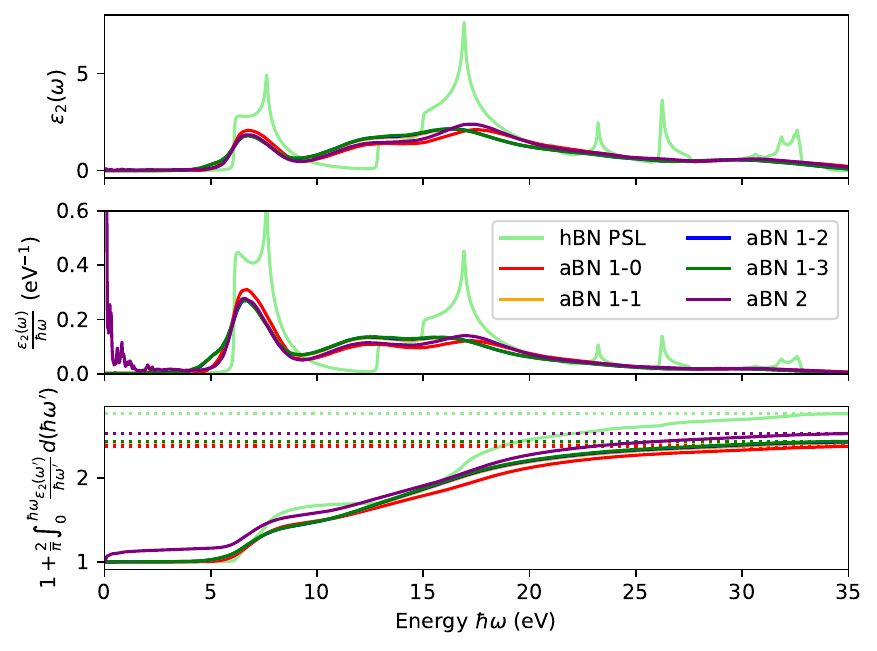}
    \caption{Top panel: imaginary part $\epsilon_2(\omega)$ of the dielectric function for the samples introduced in Section \ref{s:TightBinding} over the model's full transition energy range. Middle panel: Kramers-Kr\"onig contribution of each transition energy to the static dielectric constant (see equation \ref{eq:KramersKronigZero}). Bottom panel: cumulative energy integration of the former. Dotted horizontal lines: static dielectric constants of the samples (colors are matched with the corresponding solid lines). The legend is shared by all panels.}
    \label{fig:KramersKronig}
\end{figure}

In this appendix, we present the imaginary part of the dielectric function for the systems discussed in Section \ref{s:TightBinding}  and the energy-resolved contribution of transitions to their dielectric constant over the model's whole range of excitation energies. Figure \ref{fig:KramersKronig} displays these results. Its top panel, of which figure \ref{fig:Epsilon_SK} (b) is a low-energy zoom, shows $\epsilon_2(\omega)$. As discussed in the main text, it can be seen that the van-Hove singularities are strongly suppressed in aBN compared to hBN PSL. We however note in aBN the presence of a feature above the first van-Hove singularity of hBN PSL which is absent in the latter system. Energetically, the region of the first van-Hove singularity for hBN PSL corresponds to $\pi\to\pi^*$ transitions in that system. Above it, one could expect features associated to $\pi\to\sigma^*$ and $\sigma\to\pi^*$ transitions (see top center panel of figure \ref{fig:TB_bands}), but these are symmetry-forbidden in hBN due to its planar geometry. In aBN, however, this symmetry is broken and the corresponding transitions would be allowed.

To better appreciate the contribution of transitions at each energy to the static dielectric constant $\epsilon_1(\omega=0)$, we display in the middle panel of figure \ref{fig:KramersKronig} the quantity $\frac{\epsilon_2(\omega)}{\hbar\omega}$. Indeed, at $\omega=0$, the Kramers-Kr\"onig relation of equation \ref{eq:KramersKronig} reads:
\begin{equation}
\label{eq:KramersKronigZero}
\epsilon_1(\omega=0) = 1+\frac{2}{\pi}\int_0^{+\infty}\frac{\epsilon_2(\omega^\prime)}{\hbar\omega^\prime} d\hbar\omega^\prime
\end{equation}
so that the area under the $\frac{\epsilon_2(\omega)}{\hbar\omega}$ curve between two energies can be interpreted as the contribution of associated transitions to the static dielectric constant. This representation confirms that the diminution of the static dielectric constant of the aBN samples compared to the crystalline reference is essentially due here to the loss in strength and/or density of optical transitions close to the crystalline features of $\epsilon_2$. In the case of aBN1-x samples, the contribution of mid-gap states is not very strong, owing mostly to their low number overall (see figure \ref{fig:DOS_IPR}). The situation is different in the case of the aBN2 sample (which is quenched faster): while the number of mid-gap states is not dramatically higher in this case, their contribution to $\epsilon_1(0)$ is significant due to the fact that they effectively close the gap, thus inducing very low energy transitions (i.e. with low $\hbar\omega$). This can also be seen on figure \ref{fig:Epsilon_SK} (a$^\prime$) as a strong increase in $\epsilon_1(\omega)$ as $\omega\to 0$; conversely, by driving the system at higher frequencies it may be possible to no longer be resonant with these states (see the denominators of equations \ref{eq:epsilonReal} and \ref{eq:KramersKronig}) and avoid the associated increase in $\epsilon_1$, as seems here to happen for $\hbar\omega \gtrsim 1-2 \text{ eV}$.

Finally, as a consistency check and an additional visualization of the above, we display in the bottom panel of figure \ref{fig:KramersKronig} the quantity $1+\frac{2}{\pi}\int_0^{\hbar\omega}\frac{\epsilon_2(\omega^\prime)}{\hbar\omega^\prime} d\hbar\omega^\prime$ (note the replacement of $+\infty$ by $\hbar\omega$ in the integral bound), which is thus the contribution to the static dielectric constant of transitions up to a cutoff energy of $\hbar\omega$. We verify in particular that $\epsilon_2(\omega)$ was indeed computed for energies $\hbar\omega$ high enough that the static value of $\epsilon_1$ is recovered over the chosen range.



\section*{References}
\bibliographystyle{iopart-num}
\bibliography{biblio}

\end{document}